\documentclass[useAMS,usenatbib]{mn2e}

\usepackage{aas_macros}
\usepackage{graphicx}
\usepackage{hyperref}
\usepackage{pdflscape}
\usepackage{pdfpages}

\newcommand{\xrt}{\textit{Swift}/XRT}
\newcommand{\bat}{\textit{Swift}/BAT}
\newcommand{\xs}{X-ray spectrum}
\newcommand{\po}{power-law}
\newcommand{\Po}{Power-law}
\newcommand{\lp}{log-parabola}

\newcommand{\bk}{broken power-law}

\newcommand{\nh}{N$_H$}
\newcommand{\ch}{$\chi_{red}^2$}

\newcommand{\tabref}[1]{$^{#1}$}

\title[X-ray spectral studies of TeV $\gamma$-ray emitting blazars]{X-ray spectral studies of TeV $\gamma$-ray emitting blazars }

\author[A. Wierzcholska, S.Wagner]{
    Alicja Wierzcholska$^{1,2}$\thanks{E-mail: alicja.wierzcholska@ifj.edu.pl}, Stefan J. Wagner$^{1}$\\
    $^{1}$Landessternwarte, Universit\"at Heidelberg, K\"onigstuhl 12, D 69117 Heidelberg, Germany \\
    $^{2}$Insitute of Nuclear Physics, Polish Academy of Sciences, ul. Radzikowskiego 152, 31-342 Krak\'{o}w, Poland\\
   }

\begin{document}

\date{Accepted .... Received ...; in original form ...}

\pagerange{\pageref{firstpage}--\pageref{lastpage}} \pubyear{2015}

\maketitle

\label{firstpage}

\begin{abstract}
This work is a summary of the X-ray spectral studies of 29 TeV $\gamma$-ray emitting blazars observed with \xrt\, especially focusing on sources for which X-ray regime allows to study the low and the high energy ends of the particle distributions function.
Variability studies require simultaneous coverage,
ideally sampling different flux states of each source.
This is achieved using X-ray observations by disentangling the high-energy 
end of the synchrotron emission and the low-energy end of the Compton emission, which are produced by
the same electron population. We focused on a sample of 29 TeV $\gamma$-ray
emitting blazars with the best signal-to-noise X-ray observations
collected with \xrt\ in the energy range of 0.3-10\,keV during  
10\,years of \xrt\ operations. \\
We investigate the X-ray spectral shapes and the effects of different
corrections for neutral hydrogen absorption and decompose 
the synchrotron and inverse Compton components.
In the case of 5 sources (3C\,66A, S5\,0716+714, W\,Comae, 4C\,+21.35 and BL\,Lacertae) a superposition of both components is observed in the X-ray band,
permitting simultaneous, time resolved studies of both ends of the electron
distribution.
The analysis of multi-epoch observations revealed that the break
energy of X-ray spectrum varies only by a small factor with flux changes.
Flux variability is more pronounced in
the synchrotron domain (high-energy end of the electron distribution)
than in the Compton domain (low energy end of the electron distribution).
The spectral shape of the Compton domain is stable,  while the
flux of the synchrotron domain is variable.
 These changes cannot be described by
simple variations of the cut-off energy, suggesting that the high-energy
end of the electron distribution is not generally well-described by cooling
only.

\end{abstract}

\begin{keywords}
Radiation mechanisms: non$-$thermal $--$ Galaxies: active $--$ BL Lacertae objects: general $--$ Galaxies: jets
\end{keywords}

\section{Introduction}\label{intro}
BL Lacertae (BL Lac) type objects as well as Flat Spectrum Radio Quasars (FSRQ) constitute a class of Active Galactic Nuclei (AGN). 
In a unified model,  these are the sources  characterized by polarized and highly variable non-thermal continuum emission observed from the jet pointing at small angles to the line of sight \citep[e.g.,][]{begelman84}.  The radiation observed extends from radio wavelengths up to the X-ray regime, or even - in the case of the most energetic sources -  up to high and  very high energy $\gamma$-ray range \citep[e.g.][]{Wagner2009, Vercellone,  Aharonian13_0301, Abramowski2014}.
The emission observed from blazars is known to be variable at all wavelengths on different timescales down to hours or even minutes in the most extreme cases \citep[e.g.][]{wagner,2155flare,Gopal-Krishna11,saito, Liao15}. Also, flux changes in blazars are often associated with spectral variability \citep[e.g.][]{Xue06, Bottcher10, Gaur14}. 

The spectral energy distribution (SED), in $\nu$-$\nu F_{\nu}$ representation, shows in most cases two broad emission components. The  low energy peak is usually attributed to synchrotron radiation of relativistic electrons from the jet, while the origin of the high-energy peak still remains  matter of debate. 
In the most common leptonic scenarios, the high-energy bump is explained as inverse Compton radiation from  the same population of relativistic electrons scattering either the synchrotron photons \citep[SSC, Synchrotron Self Compton models; e.g.][]{Maraschi92, Kirk98} or external photons from the surroundings, e.g. accretion disc, broad lines region or dust \citep[EC, External Compton models: e.g.][]{Dermer92, Sikora94}. Alternatively, the high-energy peak can be explained in the framework of hadronic scenarios. 
In this case  the high-energy radiation is produced in processes such   a proton synchrotron emission, synchrotron and Compton emission from secondary decay products of charged pions, or $\pi_0$ decay \citep{Aharonian00, Atoyan03, Mucke13, Bottcher13}.
Additionally,  \cite{Petropoulou2015} have shown that Bethe-Heitler emission can explain SED of blazars in the case when the  typical double-humpted structure is not clearly visible.

Depending on the frequency of the low-energy peak, BL Lac type blazars can be subdivided into high-, intermediate- and low-energy peaked objects: HBL, IBL, LBL, respectively \citep[see, e.g.,][]{padovani95,fossati98,Abdo2010}. For LBL type blazars the synchrotron peak is located in the infrared regime  ($\nu_{s}\leq10^{14}$\,Hz), for IBL blazars in the optical-UV range ($10^{14}$\,Hz$<\nu_{s}\leq10^{15}$\,Hz), while for HBL blazars in the X-ray domain ($\nu_{s}>10^{15}$\,Hz) \citep{Abdo2010}. It has been proposed  and connected with decreasing bolometric luminosities and $\gamma$-ray dominance by \cite{fossati98}. According to the blazar sequence, HBL blazars are  synchrotron dominated, while FSRQs are $\gamma$-ray dominated. In the later work the concept of the blazars sequence has been updated with a large sample of radio-loud AGNs to the blazars envelope \citep{Meyer11}.

The different classes of blazars could also be distinguished according to the location of the X-ray spectrum in SED. 
For HBL type blazars the X-ray spectrum is usually located in the synchrotron domain.
While in the case of  IBL type sources the X-ray emission usually covers both synchrotron and inverse Compton component. For several IBL type sources the upturn of SED located in the X-ray domain has been reported e.g.  BL~Lacertae \citep{Tanihata00,Ravasio02,Donato05}, W~Comae \citep{Tagliaferri00, Donato05}, S5~0716+71 \citep{Cappi94,Giommi99,Tagliaferri03,Donato05,Ferrero06, Wierzcholska_s5_2015}, AO~0235+16 \citep{Raiteri06}, OQ~5310 \citep{Tagliaferri03}, 3C~66A \citep{Donato05}. LBL type objects are usually characterised with the X-ray spectrum covering the inverse Compton bump.

The X-ray observations of the brightest HBL type blazars unfolded strong variability often  with large amplitude of variations, observed on long as well as on short timescales \citep{Sembay93, Tanihata00, Sembay02, Zhang02, Zhang2005}. In several mentioned cases HBLs variability amplitudes have been also found to be correlated with the energy. Furthermore, HBL blazars are known for ``harder-when-brighter'' behaviour manifested in hardening of spectral index with the increasing flux level observed \citep[e.g.][]{Pian98, Zhang2005, Zhang06b, Abramowski2012_2155}.

The X-ray variability studied in different energy bands seem to be correlated, often with  time-lags of a few hours. Usually variations observed in lower energies have lags behind those at higher energies (so-called soft time-lags), but the opposite (so-called hard time-lags) are also observed \citep[e.g.][]{Zhang06b,Fossati2000}.
Furthermore, the comparison of X-ray spectral index ($\alpha$) and flux intensity (I) presented in a $\alpha$-I plots reveals clockwise (for soft time-lags) or counter-clockwise (for hard time-lags) patterns \citep[e.g.][]{Takahashi96, Ravasio04, Brinkmann05}. The time-lags are directly related with changes of acceleration and cooling time scales \citep{Kirk98}.

The X-ray variability observed in the case of IBL and LBL type objects show similar behaviour according  to the variability time scales and amplitudes observed for HBL type blazars.  But, for these classes of blazars, as opposed to HBL type ones, the variability amplitudes seem to be anticorrelated with the emission energy \citep[e.g.][]{Giommi99, Ravasio02}. 
Furthermore, in mentioned studies no time-lags between  the low and the high energy X-ray emission has been observed. 

Previous studies focusing on the nature of the X-ray spectra disclosed its complex nature and several authors discussed features of blazars behaviour in this domain.
\cite{Worrall90} analysed  large sample of BL Lac type blazars observed with \textit{Einstein} which were well described with the single \po\ model with a wide range of spectral indices. This model as well as  the \bk\ one were also successfully used to describe the X-ray spectra in the later works for data collected with  different  instruments like  ASCA \citep[e.g.][]{Kubo98}, \textit{Beppo}SAX \citep[e.g.][]{Wolter98,Padovani01, Padovani04}, HEAO-1 \citep[e.g.][]{Sambruna94} and ROSAT observations \citep[e.g.][]{Perlman96b, Urry96, Sambruna96}.

Alternatively, in a several works  the X-ray spectra are well described with the curved \lp\ model.
\cite{Inoue96, Tavecchio98, Giommi02, Donato05} studied the X-ray observations of blazars performed with \textit{Beppo}SAX and  found this model as preferable for at least 50$\%$ of the sources analysed. 
In the later studies   \cite{Massaro04b, Massaro04a, Massaro06} analysed the X-ray spectra for Mrk\,421 and Mrk\,501  and found the \lp\ as the best model for characterization of their spectra in different activity states. The authors also suggested a possible interpretation of the mentioned model in term of statistical  particle acceleration with the assumption that the probability of the increase of particles energy is a decreasing function of an energy. 
Furthermore \cite{Massaro08} enlarged their sample to 15 objects and found the \lp\ as the best model for most of the blazars spectra studied.

The mentioned studies revealed that the X-ray spectra are characterized with a wide range of spectral indices.
Following the distributions presented by \cite{Donato05}, which are consistent with the previous and the later studies by e.g. \cite{Perlman05, Tramacere07, Massaro08}, the X-ray spectral indices of HBL type blazars are between $\Gamma_X=1.8$ and $\Gamma_X=3.0$, for LBL type objects this quantity is between  $\Gamma_X=1.5$ and $\Gamma_X=1.8$, while in the case of FSRQs, spectra are best described with the spectral indices between $\Gamma_X=1.0$ and $\Gamma_X=2.0$.

The motivation of this work is to test if the X-ray spectra observed with \xrt\ in the energy range of 0.3-10\,keV are well described with the single \po\ model or the continuously curved \lp. Both characteristics are reasonable and widely discussed in the context of models used for description of synchrotron ageing and acceleration processes \citep[e.g.][]{Leahy91, Massaro04b, Massaro04a}. Also, we would like to check for how many of selected TeV emitting blazar the X-ray regime is a place, where the low- and the high-energy spectral component meet or if the X-ray spectral shape allows us to predict where this point for particular blazars is located in broadband SED. 
Finally, different values of the Galactic column absorption provided in several surveys \citep[][]{Dickey90, Kalberla05, Willingale13} give us reason to study, which values are preferable and how changes in the Galactic column density values influence the X-ray spectral properties.

The paper is organized as follows: the Swift Gamma-Ray Burst Mission is described in Sect.~\ref{swift}, while Sect.~\ref{sample} gives information about the sample selection and data analysis. All selected sources are described in Sect.~\ref{catalogue}. Sect.~\ref{variabilitysl} focusses on 5 blazars, for which an upturn point is visible in the X-ray regime. The work is summarized in Sect.~\ref{discussion}.

\section{The Swift Gamma-Ray Burst Mission}\label{swift}
The Swift Gamma-Ray Burst Mission (hereafter \textit{Swift})  \citep{Gehrels04}, launched in November 2004, is a multi-wavelength space observatory, equipped with three instruments: the Burst Alert Telescope \citep[BAT,][]{Barthelmy05}, the X-ray Telescope \citep[XRT,][]{Burrows05} and the Ultraviolet/Optical Telescope \citep[UVOT,][]{Roming05}.  \xrt\ operates in the energy range of  0.3-10\,keV.

The XRT detector can be operated in different modes: the Windowed Timing (WT), the Photon Counting (PC) and the Imaging (IM); the forth - Photo-Diode mode - is not working since May 2005. In the PC mode the full CCD chip is read out every 2.5 seconds and it is used for sources with count rate smaller than 1\,count per second. Observations provided in this mode have full imaging and spectroscopic resolution.
The WT mode is characterized by 1.7\,ms time resolution and full energy resolution. The mode is usually used to observe the sources with count rate larger than 1\,count per second. During the monitoring of a chosen source, the detector can switch between the PC and the WT mode, to optimize the observations.
The third, Imaging mode is used only to obtain the first X-ray position of a newly detected Gamma Ray Burst. 

\section{Sample selection and data analysis}\label{sample}
The catalogue of TeV sources (TeVCat\footnote{\url{http://tevcat.uchicago.edu/}}) includes 161\footnote{The number of objects collected in catalogue up to June 2015.} objects and 58 of them are classified as blazars. All of the blazars were observed several times with \xrt. The main properties of the TeV blazars, as well as the total exposure in the PC mode and integrated flux for the energy range of 0.3-10\,keV obtained using the online analysis of \xrt\ data \citep{Evans09} are collected in Table\,\ref{table_tevblazars}. Since the aim of this work is to focus on properties of the average  X-ray spectra for particular blazars,  the further analysis has been limited to objects with the best \xrt\ data  for which
the product of the integrated flux and the exposure is greater than 3.0\,$\cdot$10$^{-10}$\,erg\,cm$^{-2}$. 

This allowed to single out 29 objects with high signal-to-noise level, and for which \xrt\ data allow to obtain good-quality spectra. The sources chosen are marked in the last column in Table\,\ref{table_tevblazars}. One object, namely, IC\,310, fullfilled our criteria but it is not included in our sample, since in the literature it is not classified as a blazar, but as a radio-galaxy or a transition object between radio galaxy and BL Lac type source \citep[e.g.][]{Kadler_2012, Aleksic_ic310_2014}.

This study is limited to  observations taken in the PC mode, since data taken in the WT mode are affected by charge redistribution problems inherently related to the way in which the CCD of the instrument is read. These depend on the position of the source on the CCD before
binning the read out on a 10\,pixel basis and makes the response position-dependent.
The effect is small for mildly absorbed sources, but in this study we focus on long time integrated spectra, so  a number of photons is huge and it can be seen as an  impact on the residuals in spectra.\footnote{Private communication by Sergio Campana.}
The effect was also noticed in \xrt\ observation in WT mode by e.g. \cite{Massaro08}. 

For 29 selected blazars \xrt\ data were analysed using HEASoft package v.\,6.16 software\footnote{\url{http://heasarc.gsfc.nasa.gov/docs/software/lheasoft}} with CALDB v.\,20140120. 
All events are cleaned and calibrated  using the \verb|xrtpipeline| task. Data in the energy range of 0.3-10\,keV with grades 0-12 are analysed. 
For the spectral studies, data are grouped using the \verb|grappha| tool to have a minimum 20 counts per bin  and the co-added spectra are fitted using XSPEC v.\,12.8.2 \citep{Arnaud96}.

In the first step two models were used to fit \xrt\ spectrum:

\begin{itemize}
 \item a single power-law defined as
 \begin{equation}
\frac{dN}{dE}=N_p  \left( \frac{E}{E_0}\right)^{-{\Gamma}},
\end{equation}
 with the Galactic absorption,
\item a logarithmic parabola defined as
 \begin{equation}
\frac{dN}{dE}=N_l  \left( \frac{E}{E_0}\right)^{-({\alpha+\beta \log (E/E_0)})},
\end{equation}
 with the Galactic absorption.
 \end{itemize}
Here, the \po\ model is characterized by the normalization $N_p$ and the spectral index $\Gamma$ (at energy  $E_0$), while the \lp\ model is described with the normalization $N_l$ and the curvature parameter $\beta$ and the spectral index $\alpha$ (at energy  $E_0$). 
In both cases the scale energy $E_0$ is fixed at 1\,keV. 
In each case  four different values of the Galactic absorption parameter are tested:

\begin{itemize}
\item N$_{H}^{LAB}$ taken from the Leiden Argentine Bonn Survey \citep[LAB,][]{Kalberla05}. The survey gives an information about N$_{H}^{LAB}$ values for different region of the sky.  
The map was obtained by merging two surveys: the  Leiden/Dwingeloo  Survey \citep{Hartmann97} and Instituto Argentino de Radioastronoma Survey \citep{Arna00, Bajaja05}.
The beamsize is 36\,\textit{arcmin} for declinations $>$ $-27.5 \deg$ and 30\,\textit{arcmin} for declinations $<$ $-27.5 \deg$. The uncertainties are in each case at a 2-3$\%$ level.

\item N$_{H}^{DL}$ taken from the survey by \citep{Dickey90} (hereafter DL). The map was obtained by merging several surveys and averaging into $1x1$ degree bins. 

\item N$_{H}^{Will}$ taken form the survey by \cite{Willingale13}  (hereafter Will). The survey gives  values of N$_{H,tot}$, which includes both the atomic gas column density N$_{HI}$ and  the molecular hydrogen column density N$_{H_2}$.  N$_{HI}$ is taken form LAB survey, while
N$_{H_2}$ is estimated using the maps of dust infrared emission by \cite{Schlegel98} and the dust-gas ratio by \cite{Dame01}.

\item free N$_H^{free}$ value.
\end{itemize}
All \nh\ values are collected in Table~\ref{table_nh} (in the online material).
Additionally, in the case when the \lp\ model is the preferable one and  $\beta$ in this model is consistently negative for all \nh\ values tested
the spectrum is also fitted with the  \bk\ model (also with the Galactic absorption) described as:
\begin{equation}
\frac{dN}{dE} = N_b \times\left\{\begin{array}{ll} (E/E_b)^{-\Gamma_1} & \mbox{if $E < E_b$,}\\ (E/E_b)^{-\Gamma_2} & \mbox{otherwise.} \end{array}\right. 
\end{equation}
This model is characterized by normalization $N_b$,  two spectral indices $\Gamma_1$ and $\Gamma_2$, and the break energy $E_b$.
It allows to determine the value of the break energy, $E_{b}$, the exact point at which spectrum is broken. This parameter is important in the case of blazars for which the X-ray spectrum covers both synchrotron and inverse Compton components. 
Since BL\,Lacertae is known to have a spectral upturn located in the X-ray domain \citep[e.g.][]{Tanihata00, Ravasio02}, we also decided to fit the spectrum of this blazars with the \bk\ model. 

For  all the models mentioned the fitting method in XSPEC is set to \verb|leven|, a minimization
method using the modified Levenberg-Marquardt algorithm \citep{Marquardt63, bevington2003data}.

\section{The  \xrt\ spectral catalogue}\label{catalogue}
Following the procedure described in Sect.~\ref{sample} for 29 selected objects \xrt\ spectra has been fitted with log-parabola and power-law models. The fit parameters for the \lp\ and the \po\ fits are collected in Table~\ref{table_fitpar_lp} and Table~\ref{table_fitpar_po} (in the online material), respectively. Since the \lp\ and the \po\ are nested models, the goodness of both fitted models  is compared using the F-test \citep[e.g.][]{bevington2003data}. The test statistics value as well as the corresponding probability are collected in Table~\ref{table_fitpar_po} (in the online material).
For the additional analysis with the \bk\ model the spectral fit parameters are collected in Table~\ref{table_fitpar_bk} (in the online material).

\begin{figure*}
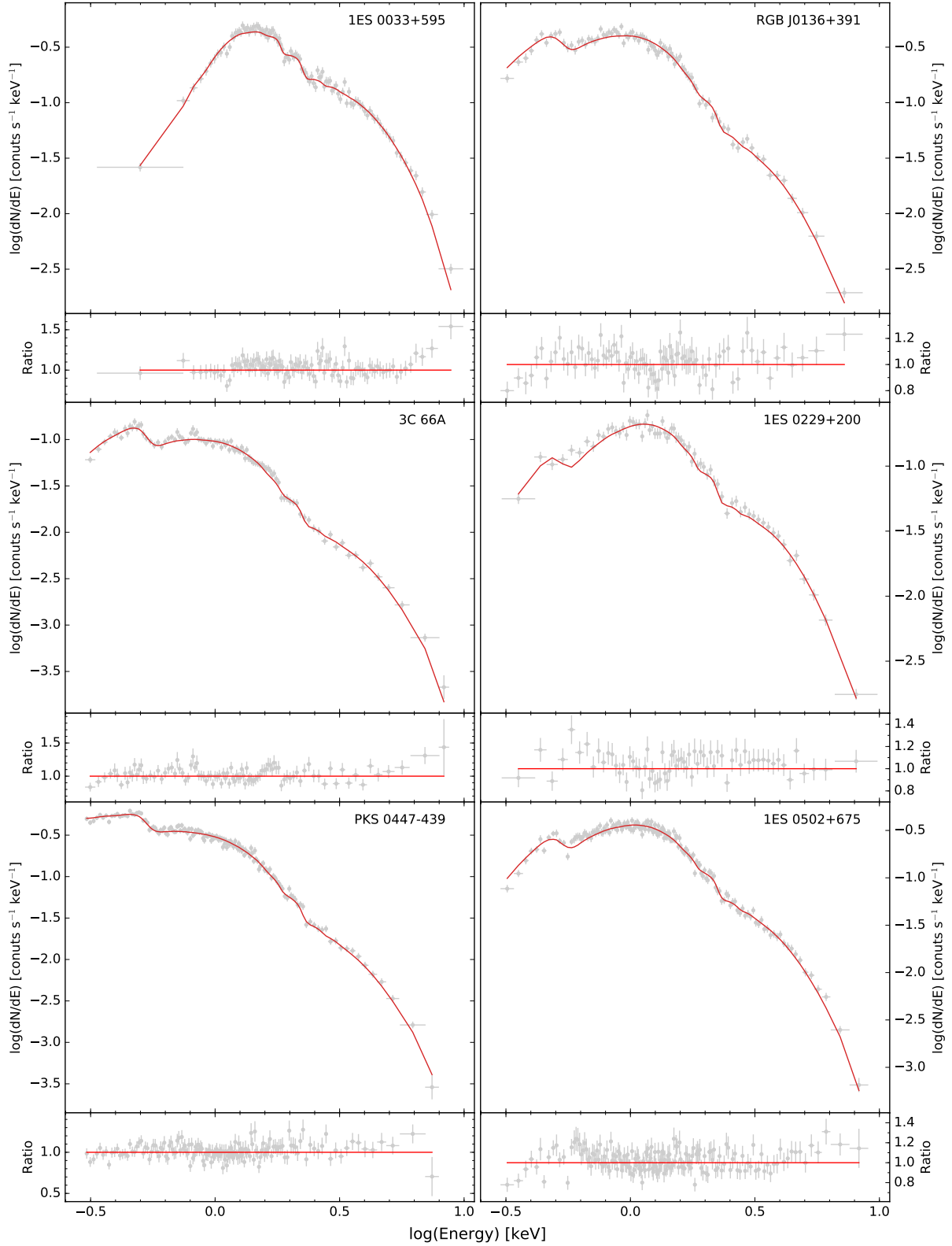

\caption{The X-ray spectra fitted with the  \lp\ model with the Galactic absorption value taken from LAB survey and the ratios (data points divided by the folded model). The  plots present spectra for: 1ES\,0033+595, RGB\,J0136+391, 3C\,66A, 1ES\,0229+200, PKS\,0447-437, 1ES\,0502+675. The name of each object is also given in the upper right corner of each plot.  }
\label{spec_1}
\end{figure*}

\begin{figure*}
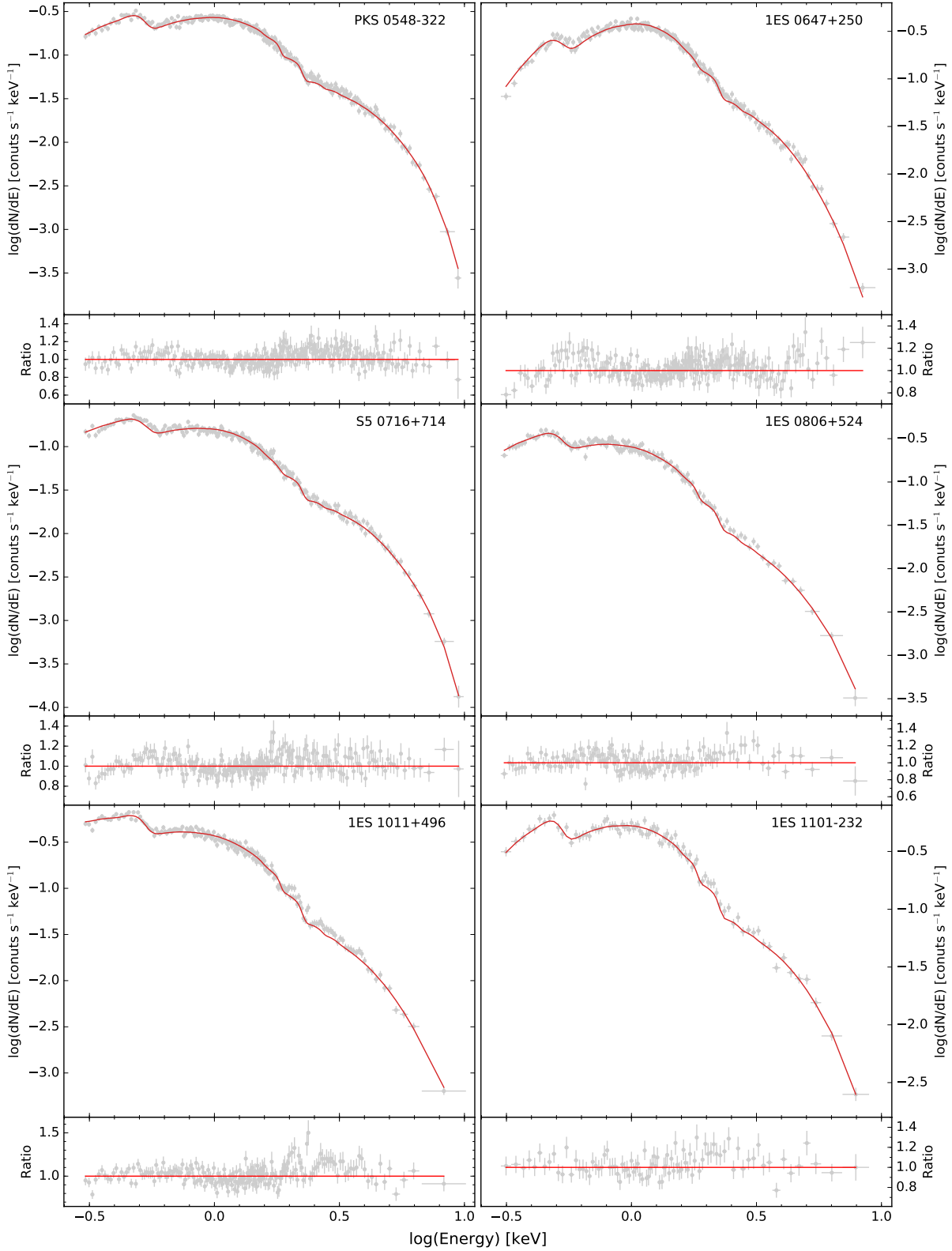

\caption{The X-ray spectra fitted with the \lp\ model with the Galactic absorption value taken from LAB survey and the ratios (data points divided by the folded model). The following plots present spectra for: PKS\,0548-322, 1ES\,0647+250, S5\,0716+714, 1ES\,0806+524, 1ES\,1011+496, 1ES\,1101-232. The name of each object is also given in the upper right corner of each plot. }
\label{spec_2}
\end{figure*}

\begin{figure*}
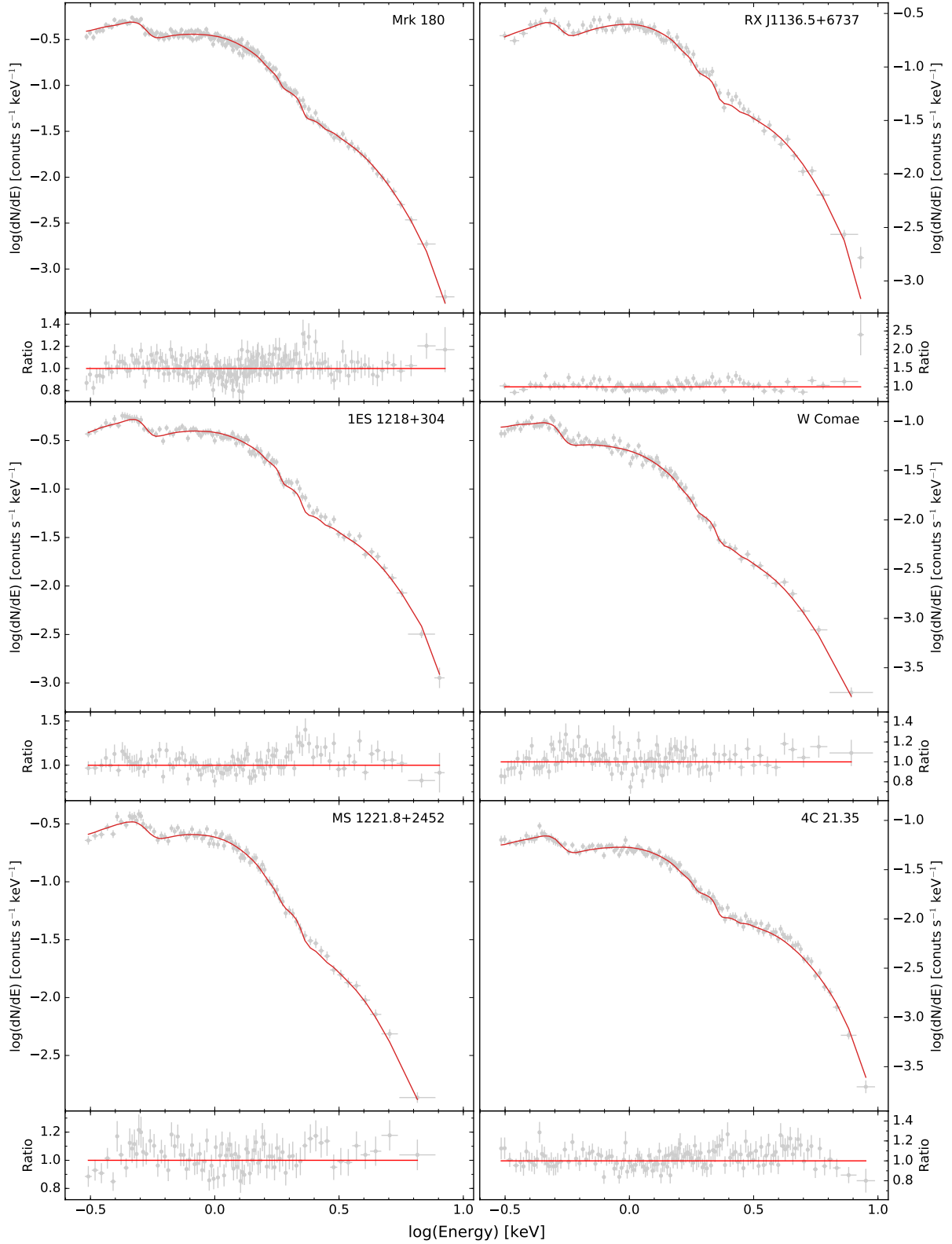

\caption{The X-ray spectra fitted with the \lp\ model with the Galactic absorption value taken from LAB survey and the ratios (data points divided by the folded model). The following plots present spectra for: Mrk\,180, RX\,J1136.5+6737, 1ES\,1218+304, W\,Comae, MS\,1221.8+2452, 4C\,21.35. The name of each object is also given in the upper right corner of each plot. }
\label{spec_3}
\end{figure*}

\begin{figure*}
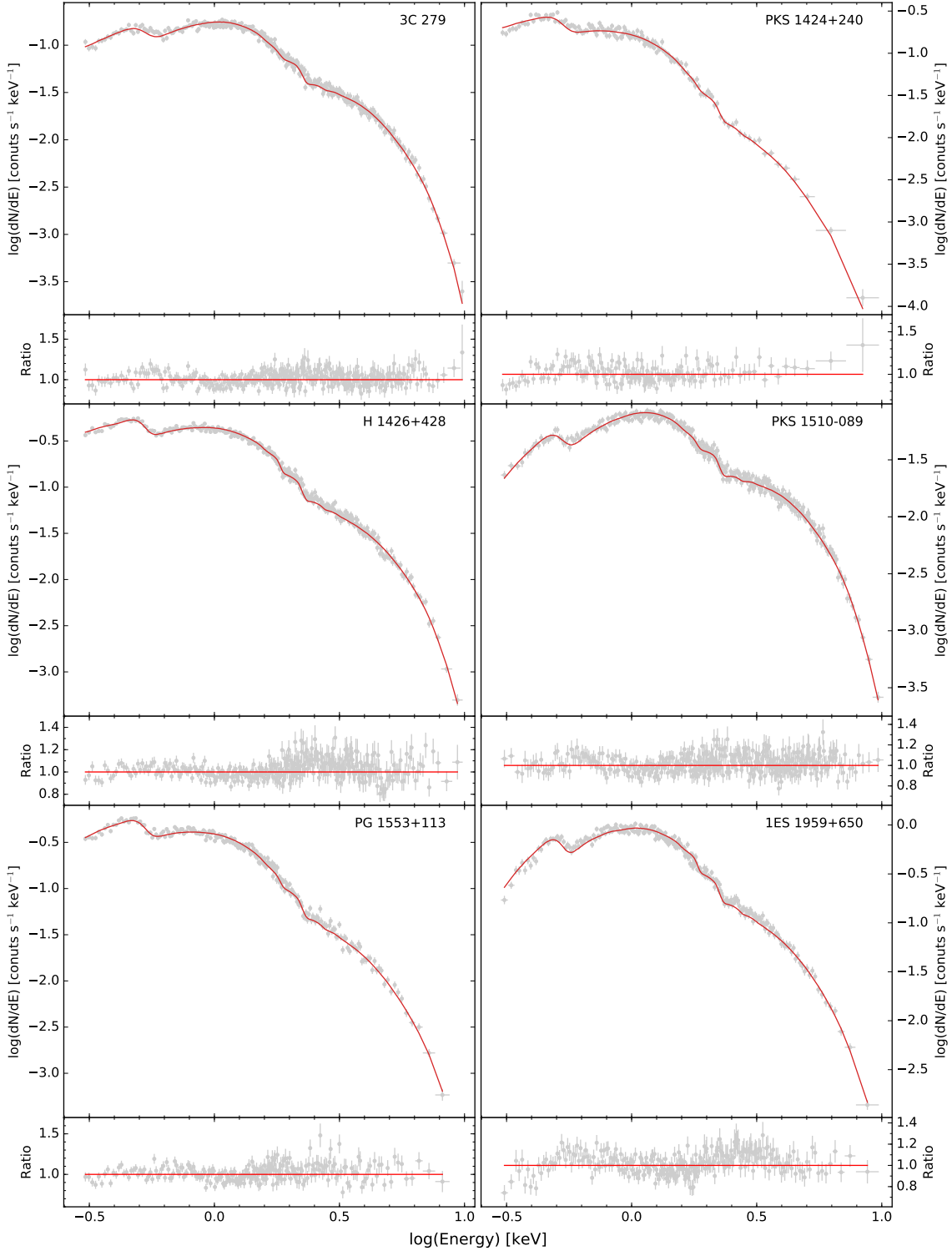

\caption{The X-ray spectra fitted with the \lp\ model with the Galactic absorption value taken from LAB survey and the ratios (data points divided by the folded model). The following plots present spectra for: 3C\,279,  PKS\,1424+240, H\,1426+428, PKS\,1510-089, PG\,1553+113, 1ES\,1959+650. The name of each object is also given in the upper right corner of each plot. }
\label{spec_4}
\end{figure*}

Spectral fits for the \lp\ and the \po\ models are presented in Fig.~\ref{seds_lp} and  Fig.~\ref{seds_po} (in the online material). Both figures presents data points fitted with  \nh\ value from the LAB survey. Four  models are presented for each blazar with colours denoting four different \nh\ values: red is used for the \nh\ value taken from \cite{Kalberla05} survey, blue for the one from \cite{Dickey90}, magenta for the \nh\ value from \cite{Willingale13} and green for the free \nh\ value. Figures  \ref{spec_1}--\ref{spec_5} (in the online material) show  spectra for all selected 29 objects and data shown in the figures are always fitted with the \lp\ model with the \nh\ value taken from the LAB survey.

Different models and \nh\ values tested lead to the conclusion that in the most cases of TeV $\gamma$-ray emitting blazars studied the preferable description of the X-ray spectrum is the \lp\ model. Only in the cases 1ES\,1101-232, Mrk\,180 and PKS\,2005-489,  the F-test values imply that the average spectrum is best fitted  with a single \po\ model.

The preferable fits do not allow us to distinguish a best, universal \nh\ value, which could be used in all cases. 
In the sample of 29 sources we found that the values taken from LAB, \cite{Willingale13}, \cite{Dickey90} are preferable in 6, 8, 5 sources, respectively, 
 while in the other 10 cases, a free \nh\ value provides the best \ch result. The selection of the preferable model and \nh\ value was based not only on the fit to the X-ray data characterised with  $\chi_{red}^2$, but also in divergent cases we investigate the shape of the fitted spectrum in the context of the overall broadband spectral energy distribution built from archival data\footnote{\url{http://www.asdc.asi.it/}}.
In the sample studied there are a few objects (such like PKS\,0548-322, H\,1426+428 or PKS 2155-304) for which   different \nh\ value  result in a very similar X-ray spectral curvature. But there are also several cases for which different surveys provide different Galactic column density values, which result in very different spectral curvatures. 
An  extreme example is  1ES\,0033+595 (see in Fig.~\ref{seds_lp}).
In several cases spectra with very different characterisations are fitted with very similar values of \ch.

In our collection of 29 X-ray spectra we found that the shape of spectra is characterised with $\alpha$ or $\Gamma$ (depending on the preferred model of the spectrum) in the range  1.3-2.6. 
For the fit with a \lp\ model the curvature parameter $\beta$ lies between $-0.100$ and $-0.347$ for the concave curved spectra and between 0.066 and 0.477 for the convex curved spectra. 
Below, main properties of the X-ray spectra for the selected 29 objects in the context  of our and previous studies are discussed: 

\subsection{Notes on individual objects}

\paragraph*{1ES\,0033$+$595:}
The \xs\ of this blazar is well described with the \lp\ model ($\chi_{red}^2\sim1.0$). The \po\ fits result in $\chi_{red}^2\sim1.8$ and it is less favoured. Comparing different values of the Galactic column density, we conclude that the best fit to data is obtained using fixed \nh\ value provided by \cite{Willingale13} or with free \nh\  value. A low value of \nh, e.g. from \cite{Kalberla05} or \cite{Dickey90}, causes an artificial bump in the synchrotron component in SED. 1ES\,0033+595 is an extreme example of a source for which different \nh\ values used result in significant changes in the spectral shapes, while $\chi_{red}^2$ values differ only slightly. 

Previous studies of the X-ray observations of 1ES\,0033$+$595 with \textit{Beppo}SAX by \cite{Giommi02} also favour the  \lp\ model for the description of the \xs, with $\alpha=1.4$ and $\beta=0.4$. These results are consistent with the ones obtained in this paper in the case of \lp\ fit with \nh\ value taken from \cite{Willingale13}.
\cite{Donato05} claimed a \po\ as a good model for later \textit{Beppo}SAX observations with spectral index of $\Gamma=2.1$ and free \nh\ value significantly smaller than \nh\ values from discussed surveys. 

\paragraph*{RGB\,J0136$+$391:}
The X-ray spectrum of the blazar is well characterized with the \lp\ model with all \nh\ values tested or with the \po\ with free \nh\ value and the one provided by \cite{Willingale13}. The preferable fits result in \ch\ of about 0.9, while in the latter case \ch\ is about 1.3.
\cite{Donato01} found single \po\ with free \nh\ value as the favourable model for this source.

\paragraph*{3C\,66A:} 
The X-ray spectrum for the blazar is best described with the \lp\ model ($\chi_{red}^2\sim$0.95). For the comparison the \po\ model fitted to data result in $\chi_{red}^2\sim$1.1-1.2.
 Negative values of the $\beta$ parameter in all \lp\ fits confirm the concave curvature of the fitted model and indicate that the turning point of SED is located in the \xrt\ energy range. We found the break energy value of (3.4 $\pm$ 0.3)\,keV by fitting also the \bk\ model to data. 
  We also found \nh\ value from \cite{Kalberla05} as the best one for the blazar.
The \lp\ fit with free value of \nh\ results in parameter value significantly smaller than the ones provided by the surveys considered. 

In the previous studies \cite{Donato05}  found \po\ with free \nh\ value or curved \po\ as preferable description of \xs. We note that in the latter case the curvature is not significant. The spectral indices mentioned by the authors ($\Gamma=2.2$-2.4) is about 5$\%$ smaller than results presented in this paper.

\paragraph*{1ES\,0229$+$200:}
In the case of this blazar the comparison of the \lp\ and the \po\ fits slightly favours description of the spectrum in the X-ray domain with the first ones. The \lp\ model result in $\chi_{red}^2\sim1.16$, while the \po\ one results in $\chi_{red}^2\sim1.20$-1.38. Regarding the different values of the Galactic column density tested, the  value  taken from \cite{Kalberla05} survey is preferred. 

In the previous studies the curvature of the spectrum in the X-ray domain was also suggested by \cite{Massaro08}  for observations obtained with   \textit{Beppo}SAX, XMM-Newton and \xrt. The results obtained by authors are consistent with results from this paper with \nh\ value from \cite{Willingale13} survey. 
\cite{Donato05} analysed the \textit{Beppo}SAX observations of the blazar and found the \po\ model with free \nh\ value as the best one for this source. The spectral index obtained by \cite{Donato05} is $\Gamma=2.0$, and it is significantly larger than values from this paper.

\paragraph*{PKS\,0447$-$439:}
The spectrum in the X-ray range for this source is well described with the \lp\ model with \ch$\sim1.1$. For the comparison the \po\ model fitted to data results in \ch$\sim1.2$-1.4. With different values of \nh\ tested   the one provided by \cite{Dickey90} is found as the best one.

In the previous studies of the X-ray data collected with \xrt\ \cite{Abramowski13_0447} described the blazar spectrum with the \po\ model with the spectral index $\Gamma=2.8$-3.3. The authors did not find any significant improvement when using the \bk\ model.

\paragraph*{1ES\,0502$+$675:}
The X-ray spectrum for this blazar can be  equally well characterized using either the \po\ fit or the \lp\ one. For the case of the \po\ model satisfying fit parameters are obtained using free \nh\ value or the one provided by \cite{Willingale13}, and in both cases \ch$\sim1.0$-1.1. For the \lp\ fit also the same \nh\ gives the best \ch$\sim1.0$ values.  For the case of the \lp\ model fitted with  \nh\ values provided  by \cite{Kalberla05} and \cite{Dickey90} we got also acceptable fits with \ch$\sim1.1$.
Different amounts of \nh\ are compensated by different degrees of curvature $\beta$.

No need for curved model for the description of the \textit{Beppo}SAX observations was found by \cite{Donato05}. The authors favour the single \po\ description of the X-ray spectrum, with the spectral index of 2.3. This value is similar to the one from our study for the case free \nh\ value.

\paragraph*{PKS\,0548$-$322:}
The spectrum  is well described with both a \lp\ and a \po\ model.
For the \lp\ model we do not find any preferable value of \nh. In each case the spectral fits result in very similar values of the spectral parameters and a \ch\ of about 1.0. 
In the case  of the \po\ fit with a free \nh\ value or the one taken for  \cite{Willingale13}.
Fits give a \ch\ of about 1.0-1.1.

 \cite{Donato05} found that both a \po\ and a \lp\ can characterize the X-ray spectrum  with the spectral indices of about 2.3-2.4.
\cite{Giommi02} prefer the \lp\ model for  \textit{Beppo}SAX observations with spectral index of 1.5 and the curvature parameter 0.5. For the comparison in this paper the preferable spectral index is 1.7 and the curvature parameter about 0.2.

The studies of several \textit{Beppo}SAX, XMM-Newton and \xrt\ observations of the source performed by \cite{Massaro08} confirmed that the X-ray spectrum has  a curved  log-parabolic shape.
Also, \cite{Aharonian10_0548} found the \bk\ model with spectral indices of 1.7 and 2.0 and the break energy of 1.7\,keV as better than the \po\ one for the case of \xrt\ observations of the blazar.

\paragraph*{1ES\,0647$+$250:}
The source is well described with the \lp\ model. Different values of \nh\ results in similar  \ch\ value of about 1.1, while in the case of the \po\ model \ch\ is in the range of 1.2-2.0, depending on \nh\ value used.
 1ES\,0647$+$250 is an example of a blazar, for which in the case of the \lp\ fit very similar \ch\ values for different \nh\ tested reveal significantly different curvatures with $\beta$ in the range from 0.17 up to 0.45. 

\paragraph*{S5\,0716$+$714:}
The \xrt\ spectrum  is well described by a  \lp\ model with \ch\ of about 1.0. For the comparison the single \po\ model tested gives \ch\ of about 1.2. Different values of \nh\  in the case of the \lp\ fit show slightly similar concave curvature with $\beta$ of about ($-$0.22)-($-$0.15).
Also the \bk\ one gives a reasonable fit with  \ch\ of 1.2.   In this case 
the break energy is found as  $E_{b}=(1.10\pm0.10)$\,keV. 

The curvature of the spectrum in the X-ray domain has been suggested in several studies. 
\cite{Tagliaferri03}, \cite{Ferrero06} and \cite{Donato05} found a \bk\ model as the preferable one with energy break of 1.5\,keV, 1.5-5.3\,keV and 1.75-2.73\,keV respectively.

\paragraph*{1ES\,0806$+$524:}
The best characterisation of the shape of the X-ray spectrum for this blazar is given using either the \lp\ fit, with any of tested \nh\ values, or with the \po\ model,  with a free \nh\ value. For all the \lp\ fits and the single \po\ with free \nh, \ch\ is around 1.0. 
Previous studies focusing on the \xrt\ observations of 1ES\,0806$+$524 did not reveal a need for a curved model  \citep{Acciari09_0806}. 
The authors describe source with the \po\ model with spectral index of 2.5-2.7, which is consistent with the results from this paper. 

\paragraph*{1ES\,1011$+$496:} 
The spectrum of 1ES\,1011$+$496 can be well characterized using the \lp\ model with any of considered \nh\  values. The models fitted result in \ch\ of about 0.9-1.0. For the single \po\ model \ch\ is in the range of 1.3 and 1.6, depending on \nh\ value chosen.
The previous studies of the \xrt\ observations by \cite{Reinthal12} show data that are well fitted with the \po\ model.

\paragraph*{1ES\,1101$-$232:} 
The X-ray spectrum  is well described using the \po\ as well as the  \lp\ models, in both cases with any of \nh\  values tested. In all cases \ch\ for each  of the \nh\ value tested is about 1.0. In the case of this blazar values of \nh\ taken from different surveys result in very similar fit parameters ($\alpha$, $\beta$, $\Gamma$).
 
\cite{Donato05} found a \lp\ as the preferred model for the \xs\ obtained using  \textit{Beppo}SAX observations. Also \cite{Massaro08} favour the curved model, characterized with $\alpha=1.6$-1.9 and $\beta=0.3$. 

\paragraph*{Mrk\,180:}
All spectral fits with a \lp\ model with any of \nh\ values tested here result in reasonable \ch\ value. 
 A good fit to data is obtained with a \po\ model, but only with free \nh\ value(\ch$\sim1.0$). 
The previous observation of the source obtained with \textit{Beppo}SAX are well described with the \lp\ model with $\alpha=2.1$ and $\beta=0.4$.
\citep{Giommi02}.

\paragraph*{RX\,J1136.5$+$6737:}
The \po\ model is a good description of the X-ray spectrum of this blazar (\ch\ $\sim$ 1.0). This spectrum is also well characterized with the \lp\ model. In the first case \ch\ is about 1.0, while in the latter 1.1. No preferable \nh\ value has been found for this source either in the case of 
single \po\ model fitted or the \lp\ one. 

\paragraph*{1ES\,1218$+$304:}
The \lp\ model result in similar values of  \ch$\sim1.2$ for the different values of \nh\ tested. Also, in the case of the \po\, different values of \nh\ tested give comparable results \ch$\sim1.3$. In both cases, the curvature and the spectral parameters of the fits do not differ significantly.

The \lp\ model is the favourable one for the spectrum characterization for  \textit{Beppo}SAX  \citep{Donato05, Massaro08},  XMM-Newton \citep{Massaro08} and \xrt\ observations \citep{Tramacere07, Massaro08}. Furthermore \cite{Giommi02} found that \textit{Beppo}SAX observations in the X-ray range can be well described with a sum of two \po\ models.

\paragraph*{W\,Comae:}
The \lp\ model is a more plausible scenario for the X-ray spectrum characterization (\ch\ of about 1.0). 
Since W\,Comae is fitted with concave curvature, in order to find the  break energy for this source, we also fitted the X-ray observations using the \bk\ model. We found $E_{b} = (2.01 \pm 0.22)$\,keV.
In the case of the \lp\ model fitted the comparison of different \nh\ values favours the \nh\ value taken from \cite{Willingale13} survey.

In previous studies the curvature of the spectrum obtained with \textit{Beppo}SAX observations was also preferred by \cite{Tagliaferri00} and \cite{Donato05}.  
\cite{Donato05} claim that the break energy is equal to $E_{b}=3.09^{+1.03}_{-1.38}$\,keV or $E_{b}=2.70^{+0.47}_{-1.33}$\,keV. \cite{Tagliaferri00} has found similar values of $E_{b}$.

\paragraph*{MS\,1221.8$+$2452}
The blazar in the X-ray regime is described with the \lp\ model  with any of \nh\ values tested and with the \po\ model with free value of the Galactic absorption column density. For  the other cases \ch\ is about 1.5, while for the preferable cases \ch$\,\sim 1.1$.

\paragraph*{4C\,$+$21.35:} 
\Po\ and  \lp\ fits result in   \ch\ of about 1.0 and $\sim1.7$, respectively.

Negative values of a  $\beta$ parameter in the case of the \lp\ model imply an up-turn of SED located in the X-ray range.
The broken \po\ fit to data placed the break energy at $(1.18 \pm 0.10)$\,keV. The best description of the \lp\ fits is given with free \nh\ value, but this  is much smaller than the ones provided by \cite{Kalberla05}, \cite{Dickey90} or \cite{Willingale13}. The latter discussed values of \nh\ describe the spectrum with a comparable accuracy.

\paragraph*{3C\,279:}
The X-ray spectrum is well explained using the \lp\ model and
different values of \nh\ result in consistent  \ch$\sim1.1$.  The \po\  model can also be used for this source, but only with free \nh\ value. Fits with the Galactic absorption values taken from the surveys results in \ch$\sim1.3$. For the \lp\ model different values of \nh\ tested result in very similar values of $\alpha$ and $\beta$ parameters.

In the previous studies \cite{Donato05} found that \textit{Beppo}SAX observations of 3C\,279 can be characterised with the \po\ and the \bk\ models, depending on the observation date. In the latter case the break energy point is in low energies $\sim$0.5-0.7\,keV.

\paragraph*{PKS\,1424$+$240:}
In the energy range of  0.3-10\,keV the spectrum can be well characterized with the \lp\ model with all \nh\ values tested  and with the \po\ model, but only with free \nh\ value.
In all the mentioned cases the spectral fitting results in \ch$\,\sim 1.0$. For the comparison the other \po\ fits, with \nh\ values provided by \cite{Kalberla05}, \cite{Dickey90} or \cite{Willingale13} result in \ch\ of about 1.4-1.5.
\cite{Acciari10_1424} found single \po\ model as preferable to describe X-ray spectrum of the blazar. Similar result found \cite{Aleksic_1424_2} for the \xrt\ observations of PKS\,1424$+$240. 

\paragraph*{H\,1426$+$428:}
A \lp\ model provides \ch\ value of about 1.2, while in the case of the \po\ model the fitted \ch\ is about 1.6-1.7.
All \nh\ values tested in the case of the \lp\ model give comparable results regarding the spectral index and the curvature parameter.

The curved shape of the spectrum of the blazar is also suggested in previous studies for \textit{Beppo}SAX, XMM-Newton and \xrt\ observations by  \cite{Massaro08} and \cite{Tramacere07}, while \cite{Giommi02} favours the \po\ model as a description of the \textit{Beppo}SAX observations. 

\paragraph*{PKS\,1510$-$089:} 
The \xrt\ spectrum of this blazar is best described with either   a \po\ or a  \lp\ fit. The favourable \nh\ values are the ones provided by \cite{Kalberla05} and free \nh\ value. For the \lp\ model also the \nh\ value taken from \cite{Dickey90} gives satisfying result. The free \nh\ values are smaller than the values provided in the quoted surveys.
\ch\ values of the preferred fits are close to 1.0.

\cite{Donato05} found  the \bk\ as the best description of \textit{Beppo}SAX observations. Again, in this case, the break energy is found to be in the  low energy part of  the X-ray range i.e. $E_{b}=1.41^{+1.21}_{-0.25}$\,keV. 

\paragraph*{PG\,1553$+$113:}
The X-ray spectrum can be well described using both a \po\ and a \lp\ models, but in the first case only with the free \nh\ value.
In the case of the \lp\ models similar values of \ch$\sim1.2$ are obtained using different \nh. The \lp\ model the different values of \nh\ result in very similar values of $\alpha$ and $\beta$ parameters. 

\cite{Massaro08} and \cite{Tramacere07}  described the X-ray spectra with the curved  \lp\ models. The object was also a target of studies by \cite{Donato05}, fitted the \po\ and the \bk\ models, depending on the observation date. However \cite{Giommi02} favour the \lp\ hypothesis, also for \textit{Beppo}SAX monitoring. 
More recent multi-wavelength studies of the blazar confirmed preference of a convex curved model for \xrt\ data \citep{Aleksic14_pg}.

\paragraph*{1ES\,1959$+$650:} 
A For the case of the \lp\ model \ch\ values are around 1.2-1.3, while in the case of the \po\ model \ch\ is about 1.3 for the case of free \nh\ value and the one taken from survey by \cite{Willingale13} and 2.0 in the other cases. 
In the case of the \lp\ fit with \nh\ value from \cite{Willingale13} the $\beta$ parameter is negative, which we do not find as a preferable description of the X-ray spectrum of this blazar, especially while the observations collected in the other wavelengths are considered.

 \cite{Giommi02, Donato05, Tramacere07} used the \lp\ model and  the \bk\ one as preferred for \xs\ of this blazar. 
In the recent multi-frequency studies of the source, including also \xrt\ and RXTE-PCA data, the \po\ model is found as a satisfactory description \citep{Aliu13, Aliu14_1959}.

\paragraph*{PKS\,2005$-$489:}
The X-ray spectrum can be well characterized using the \lp\ and the \po\ models. Also all values of \nh\ tested result in comparable values of \ch$\sim1.0$. All the spectral fits give very similar values of the fit parameters. 
The negative values of $\beta$ suggest a concave curvature, but the  errors calculated for the parameter are large compared to the value. The curvature effect is caused by the high energy part of the spectrum, where only a few last spectral points rise.

In the previous works \cite{Donato05} found the \bk\ as the best model to characterize \textit{Beppo}SAX observations of PKS\,2005$-$489, while \cite{Giommi02, Tramacere07, Massaro08} suggested a  description of \textit{Beppo}SAX, XMM-Newton and \xrt\ observations with the  \lp\ model. 
Several multi-wavelength campaigns targeting the source also included the X-ray observations with different instruments, both in the low and the high states of the blazar. In most of these cases soft X-ray observations are well described with the single \po\ model 
\citep[e.g.][]{Sambruna95, Perlman99, Tagliaferri01, Rector02, Acero10_2005, Aharonian11_2005}.

\paragraph*{PKS\,2155$-$304:}
For the blazar the X-ray spectrum is well fitted using the \lp\ model. Different values of \nh\ tested result in comparable $\chi_{red}^2\sim1.1$, while in the case of the \po\ model fitted $\chi_{red}^2\sim1.2$. Again, this is the case of blazar for which different values of \nh\ tested, for the \po\ as well as the \lp\ model result in very similar values of the $\alpha$, $\beta$, and $\Gamma$ parameters.

In the earlier studies \cite{Donato05} found the curved \lp\ model as the best description of \textit{Beppo}SAX observations. The same model was successfully used by \cite{Giommi02,Tramacere07,Massaro08} for \textit{Beppo}SAX, XMM-Newton and \xrt\ observations.
The studies of XMM-Newton observations of PKS\,2155$-$304 in 2006 revealed a curved shape of the spectrum and an indication of both synchrotron and inverse Compton components revealed in data \citep{Zhang08}. The results mentioned suggest that PKS\,2155$-$304 can be a rare example of HBL type blazar for which possible up-turn point of SED is found in soft X-ray regime. 
However previous studies with  \textit{Beppo}SAX observations of the source were well  described with the single \po\ or the convex \bk\ model \citep[e.g.][]{Zhang99, Chiappetti99, Zhang02}.

Also during the  multi-wavelength studies targeting the low state of the blazar X-ray spectrum was well characterized with the \bk\ model \citep{Aharonian09_2155}. 
Detailed, longterm spectral studies, including observations of the blazar obtained during the period of 2005-2012 with XMM-Newton has been successfully described with the \lp\ model \citep{Kapanadze14}.

\paragraph*{BL\,Lacertae:}
The X-ray  spectrum of this blazars can be well fitted with the \po\ as well as with the \lp\ model, but in both cases with free value of the Galactic absorption parameter and the one provided by \cite{Willingale13}.
In the case of the X-ray spectrum of BL\,Lacertae different values of \nh\ tested, for the \po\ as well as the \lp\ models, result in diametrically different values of the spectral and curvature parameters.
 Visual inspection of data suggests the curvature in the X-ray domain, but it is difficult to quantify, due to the high dispersion of data points in low energy part of the spectrum. For the preferable fits \ch is about 1.1-1.2. 

Previous studies on this blazar's spectrum in the X-ray energy band disentangle synchrotron and inverse Compton components in this range. \cite{Tanihata00} found that for the blazar in the energy range below 1\,keV a soft and steep spectrum is dominating, in contrast to hard component otherwise.
\cite{Ravasio02} reported two periods of observations of the blazar in 1999. In June the source has a concave spectrum with a very hard component above 5-6\,keV, while in December data were well fitted with the \po. 
\cite{Donato05} found the spectrum well described with the \bk\ model with the break energy $E_{b}=0.45^{+0.39}_{-0.44}$\,keV.
Also, the observations of BL\,Lacertae with INTEGRAL/IBIS are also well described with the \bk\ model with $E_{b}=1.71^{+0.0.75}_{-0.15}$\,keV.
\cite{Raiteri09} in the multi-wavelength studies including XMM-Newton observations of the source also unveiled a concave curvature in the  X-ray regime. The curved model is also the preferable one for The \textit{Beppo}SAX observations studied by \cite{Massaro08}, while \cite{Giommi02} suggested the  single \po\ as the best description for \textit{Beppo}SAX data.

\paragraph*{B3\,2247$+$381:}
The X-ray spectrum of this blazar is well characterized with the \lp\ model with any of \nh\ values tested. In each of the mentioned cases \ch\ is about 1.0. A good description of the spectrum is also given with the \po\ model with free \nh\ value and the one provided by \cite{Willingale13}. For the other \nh\ values tested, the one provided by \cite{Kalberla05} and by \cite{Dickey90}, \ch\ values are 1.6 and 1.3, respectively.

\paragraph*{1ES\,2344+514:}
The best description of the spectrum is given with the \lp\ model and a free Galactic absorption value as well as \nh\ provided by \cite{Kalberla05} and \cite{Dickey90}. For these preferable cases \ch\ is about 1.1.
The \lp\ fit with \nh\ from \cite{Willingale13} survey results in the largest value of \ch\ of about 1.3 and negative value of the $\beta$ parameter.  As in the case of 1ES\,0414$+$009, the $\beta$ value is not consistent with the SED shape \citep[see e.g.][]{Aleksic13}. 
In the context of  the \po\ model a fit with a free value of the Galactic column density gives a satisfactory result. 

Previous studies of the X-ray spectra obtained with \xrt\ and \textit{Beppo}SAX  show that these data cover the synchrotron bump in its SED. The X-ray spectrum was usually well described with the  \po\ model \citep[e.g.][]{Giommi00, Acciari11,  Aleksic13}. However a need for a curved model was  found also for the case of \textit{Beppo}SAX X-ray data by \cite{Massaro08}.
It is also worth mentioning that 1ES\,2344+514  exhibits strong spectral  variability with the synchrotron peak shifting to higher energies during flares \citep{Giommi00}. The synchrotron peak frequency during such events is at 10\,keV or above, which classifies the source as the so-called extreme HBL.

\section{Blazars with upward-curved X-ray spectra}\label{variabilitysl}

In the further studies we focus on five blazars for which a concave curvature of the spectrum suggests that  both synchrotron and inverse Compton components are located in the X-ray regime. This gives a perfect opportunity to study spectral and temporal variability of both components contemporaneously.

\begin{figure}
\centering{\includegraphics[width=0.48\textwidth]{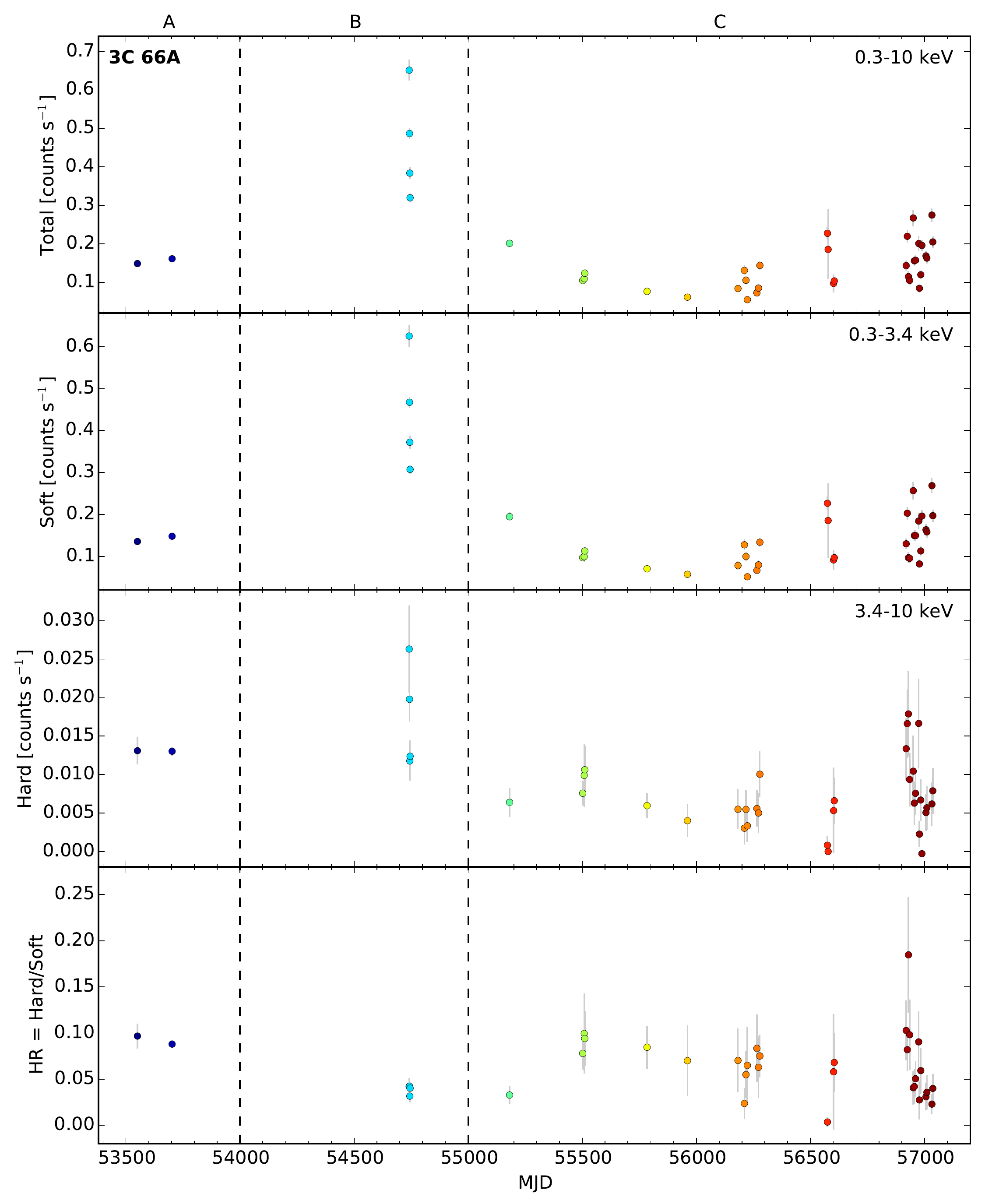}}
\caption{The longterm count rate and the hardness ratio light curve for 3C\,66A. The following panels present: counts rate in energy range of 0.3-10\,keV; the soft energy band of count flux (0.3-3.4\,keV);  the hard  energy band of count rate (3.4-10\,keV); the hardness ratio defined as ratio of hard and soft energy band count rate. The vertical dashed lines indicate selected intervals discussed in Sect.~\ref{variabilitysl}. Different colours of data points indicate the time of a given measurement. The earliest pointings are denoted by dark blue symbols and the most recent ones depicted in red. The same colour coding is used in Fig.\,\ref{flux_hr}.}
\label{lc_3c66a}
\end{figure}

\begin{figure}
\centering{\includegraphics[width=0.48\textwidth]{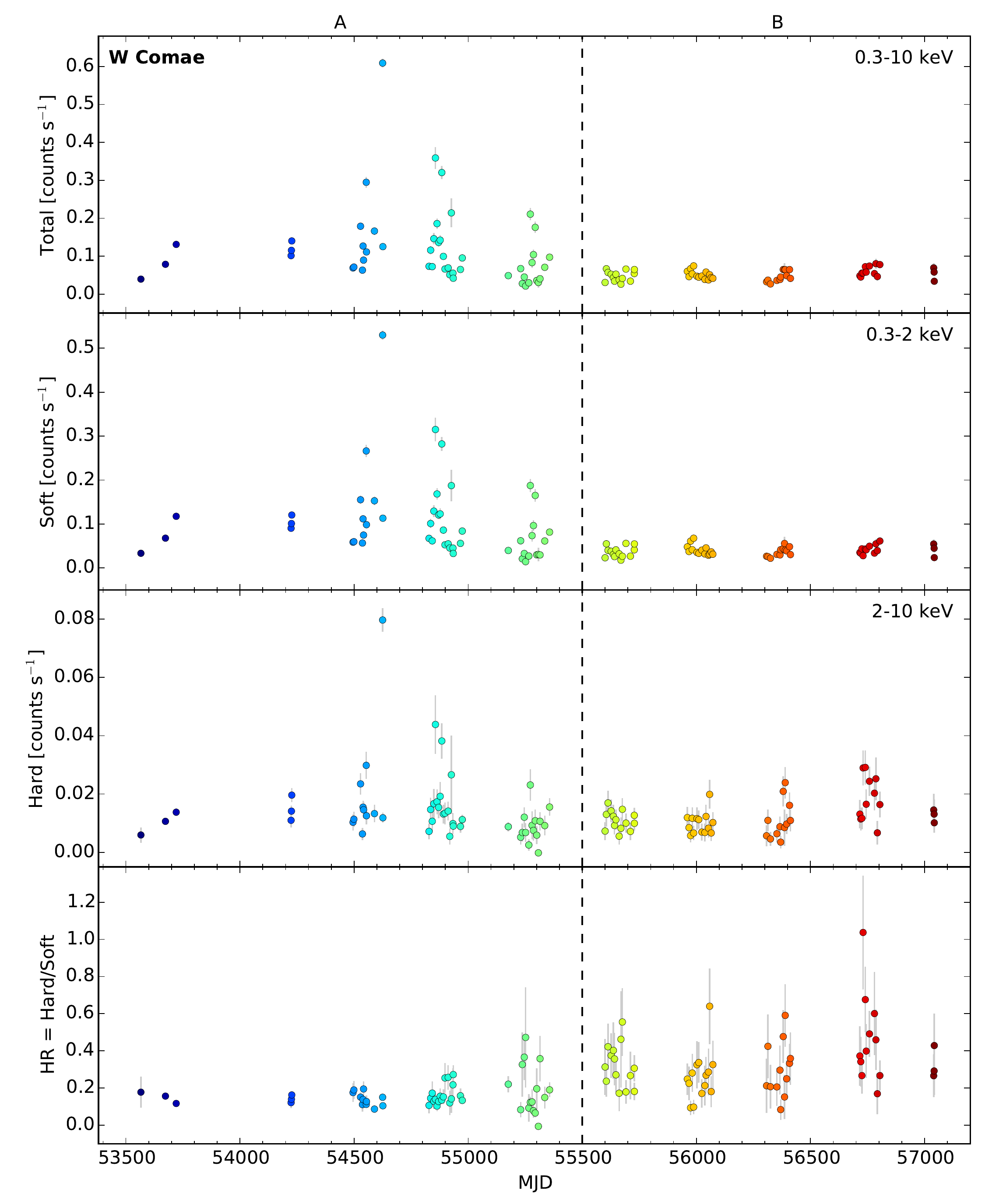}}
\caption{The longterm count rate and the hardness ratio light curve for W\,Comae. The following panels present: counts rate in energy range of 0.3-10\,keV; the soft energy band of count flux (0.3-2.0\,keV);  the hard  energy band of count rate (2.0-10\,keV); the hardness ratio defined as ratio of hard and soft energy band count rate. The vertical dashed line indicates selected intervals discussed in Sect.~\ref{variabilitysl}. Different colours of data points indicate the time of a given measurement. The earliest pointings are denoted by dark blue symbols and the most recent ones depicted in red. The same colour coding is used in Fig.\,\ref{flux_hr}.}
\label{lc_wcomea}
\end{figure}

\begin{figure}
\centering{\includegraphics[width=0.48\textwidth]{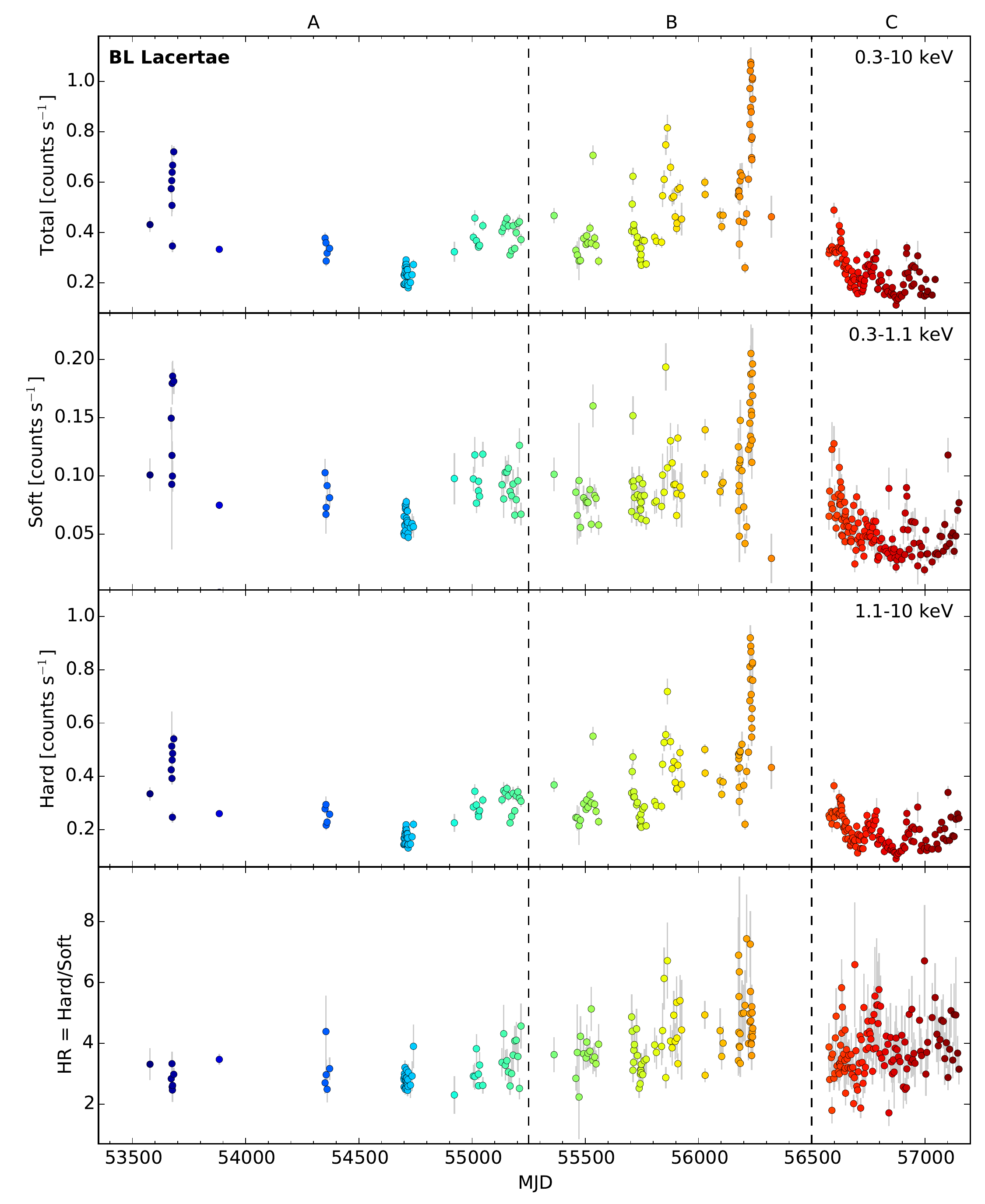}}
\caption{The longterm count rate and hardness ratio light curve for BL\,Lacertae. The following panels present: counts rate in energy range of 0.3-10\,keV; the soft energy band of count flux (0.3-1.1\,keV);  the hard  energy band of count rate (1.1-10\,keV); the hardness ratio defined as ratio of hard and soft energy band count rate. The vertical dashed lines indicate selected intervals discussed in Sect.~\ref{variabilitysl}. Different colours of data points indicate the time of a given measurement. The earliest pointings are denoted by dark blue symbols and the most recent ones depicted in red. The same colour coding is used in Fig.\,\ref{flux_hr}.  }
\label{lc_bllac}
\end{figure}

\begin{figure}
\centering{\includegraphics[width=0.48\textwidth]{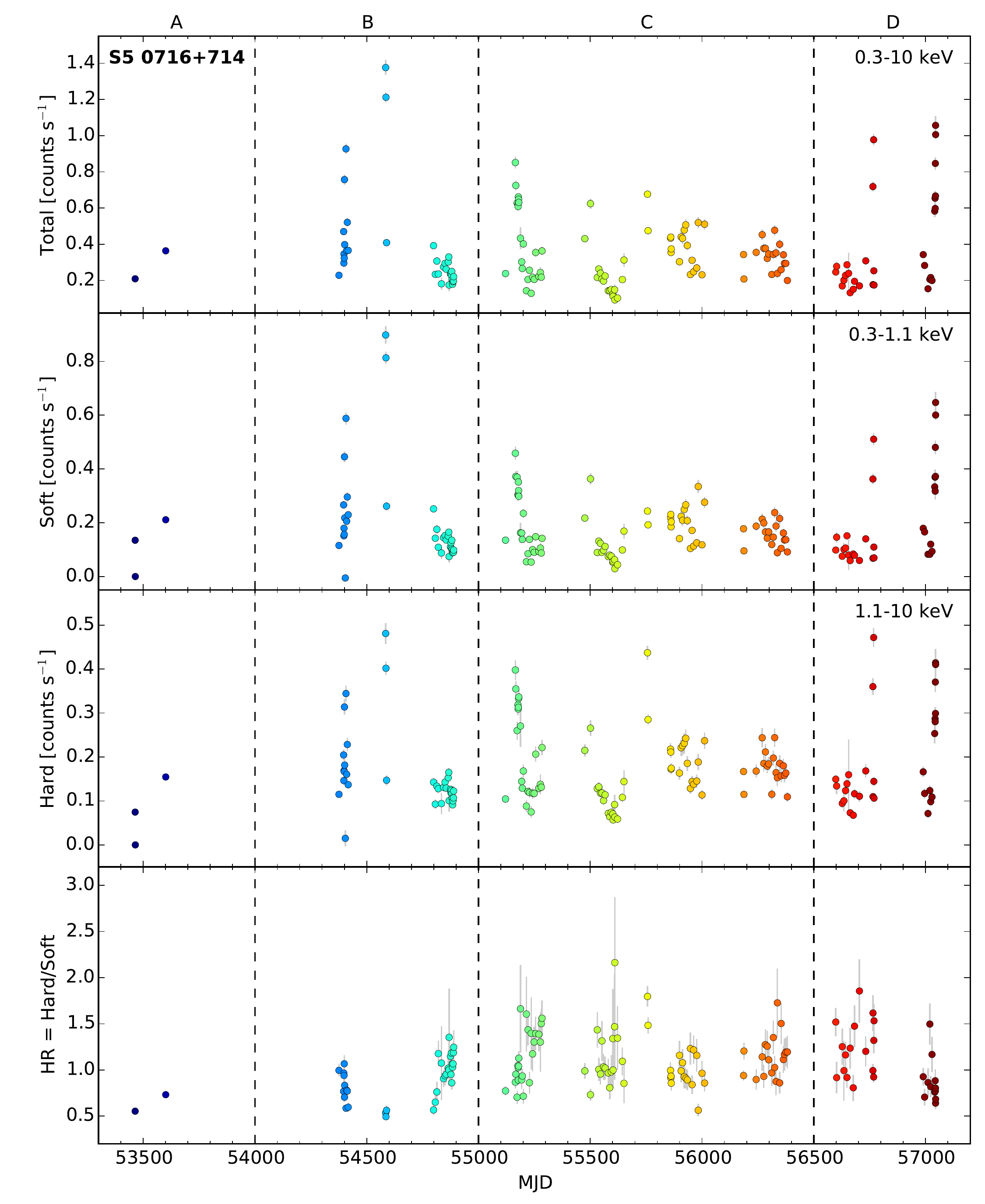}}
\caption{The longterm count rate and hardness ratio light curve for S5\,0716+714. The following panels present: counts rate in energy range of 0.3-10\,keV; the soft energy band of count flux (0.3-1.1\,keV);  the hard  energy band of count rate (1.1-10\,keV); the hardness ratio defined as ratio of hard and soft energy band count rate. The vertical dashed lines indicate selected intervals discussed in Sect.~\ref{variabilitysl}. Different colours of data points indicate the time of a given measurement. The earliest pointings are denoted by dark blue symbols and the most recent ones depicted in red. The same colour coding is used in Fig.\,\ref{flux_hr}.}
\label{lc_s50716}
\end{figure}

\begin{figure}
\centering{\includegraphics[width=0.48\textwidth]{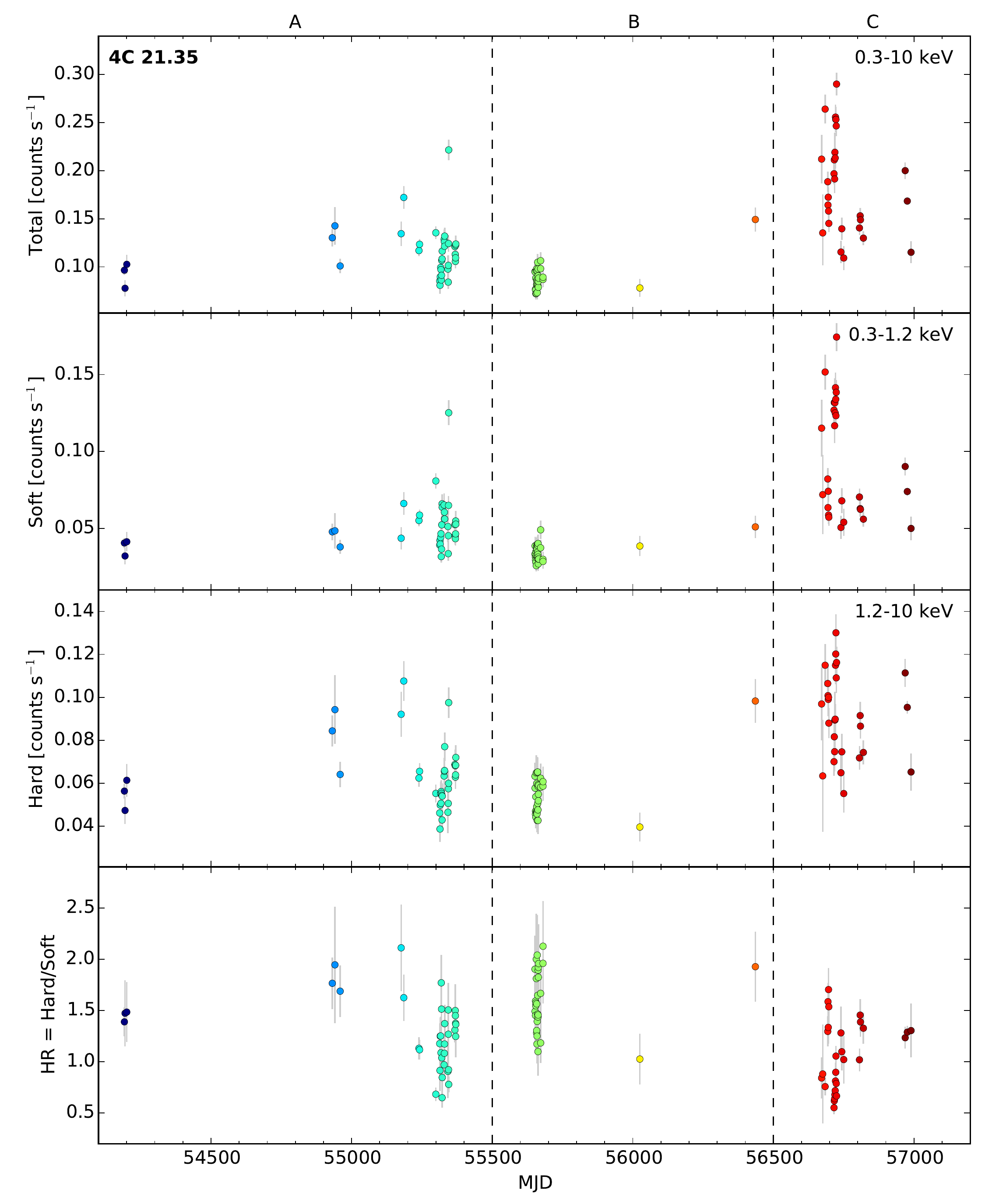}}
\caption{The longterm count rate and hardness ratio light curve for 4C\,21.35. The following panels present: counts rate in energy range of 0.3-10\,keV; the soft energy band of count flux (0.3-1.2\,keV); the hard  energy band of count rate (1.2-10\,keV); the hardness ratio defined as ratio of hard and soft energy band count rate. The vertical dashed lines indicate selected intervals discussed in Sect.~\ref{variabilitysl}. Different colours of data points indicate the time of a given measurement. The earliest pointings are denoted by dark blue symbols and the most recent ones depicted in red. The same colour coding is used in Fig.\,\ref{flux_hr}.}
\label{lc_4c2135}
\end{figure}

In the first step the longterm background-subtracted light curves for each source were inspected (upper panels of Fig.~\ref{lc_3c66a}-\ref{lc_4c2135}). In each case the presented light curves  show observations from 2005 up to October 2014 collected with \xrt\ in the PC mode in the energy range of 0.3-10\,keV. Each light curve is binned in observation-length intervals.

In order to separate synchrotron and inverse Compton components we define the upturn energy (E$_{upt}$)  being identical to the break energy value taken from the \bk\ fit, which is given in Table~\ref{table_fitpar_bk}. We define two energy bands for each source. The soft band in the energy range (0.3\,keV, $E_\textit{upt}$) and the hard energy band ($E_\textit{upt}$, 10\,keV). 
Note that $E_\textit{upt}$ is kept constant for each source throughout the 10 years period. The variations in both bands and of the hardness ratio (HR), defined as the proportion of the count rate in the  hard and the soft band, are shown in Fig.~\ref{lc_3c66a}-\ref{lc_4c2135}.

\paragraph*{3C\,66A:}
The light curve (see upper panel of Fig.~\ref{lc_3c66a}) shows variability manifested mainly in one outburst in 2008. 
The flux changes are characterized by a fractional variability amplitude \citep[][]{Edelson88}  $F_\textit{var}=(67 \pm 2) \%$ with the uncertainty calculated following \cite{Vaughan03}. Except for this outburst (with a count rate of $\sim0.7$\,count/s) the source flux level observed in different epochs  is about 0.1-0.2 counts/s. 
The soft and the hard X-ray light curves  in the energy bands of 0.3-3.4\,keV and 3.4-10\,keV, respectively are shown in Fig~\ref{lc_3c66a}. 
The fractional variability amplitudes  for the soft and hard bands are following: $F_\textit{var}=(69 \pm 2) \%$ and $F_\textit{var}=(56 \pm 1) \%$, respectively. 
The hardness ratio is characterized by a fractional variability amplitude of $F_\textit{var}=(28 \pm 1) \%$.
Hence, the most pronounced variability is observed in the soft energy band.

\begin{figure*}
\centering{\includegraphics[width=0.99\textwidth]{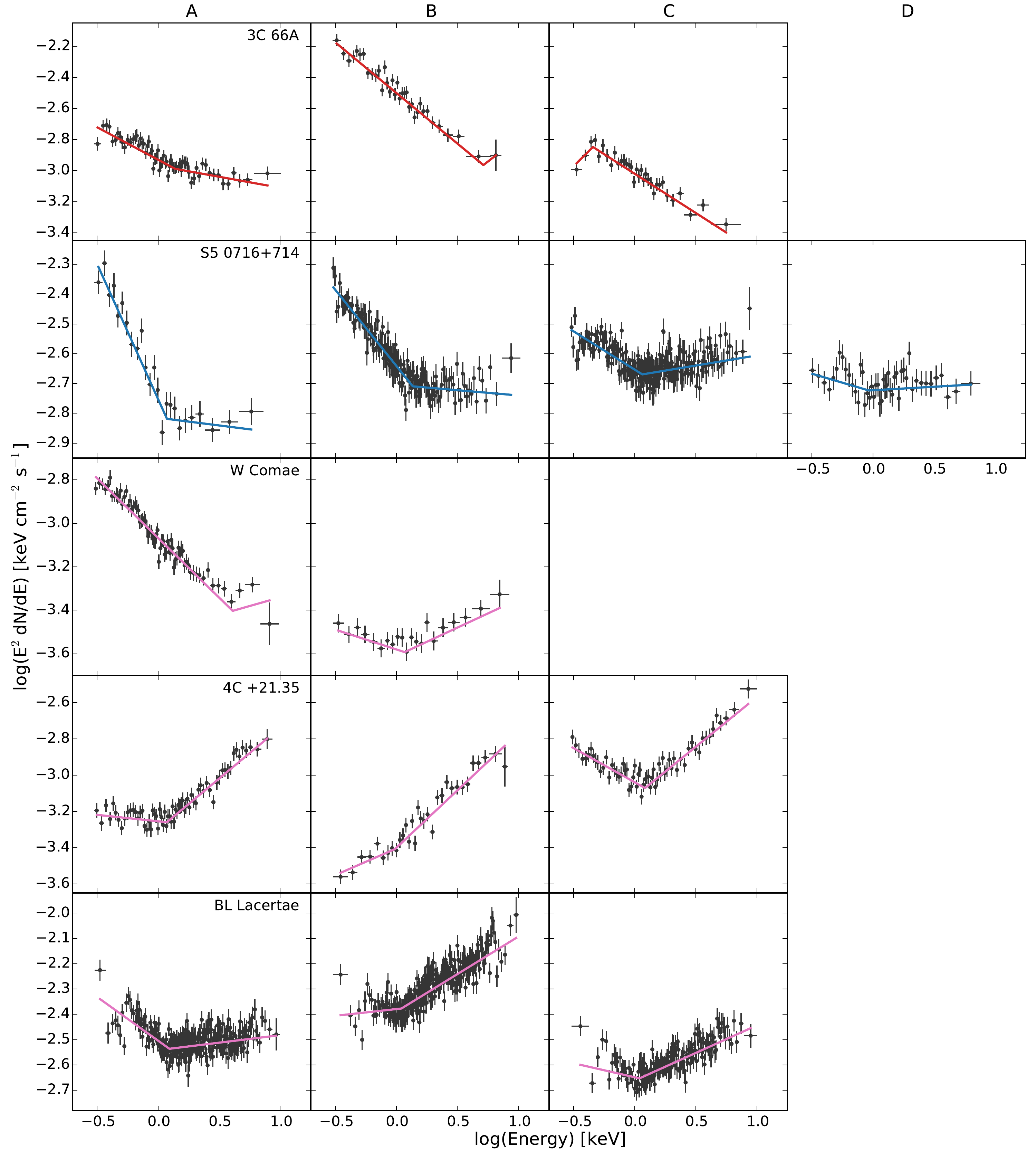}}
\caption{Spectral energy distributions for spectral fits with  \bk\ models during periods defined in Fig.~\ref{lc_3c66a}-\ref{lc_4c2135} for 3C\,66A, S5\,0716+714, W\,Comae, 4C\,21.35 and BL\,Lacertae. 
While both spectral components change in slope and flux normalization, the upturn energy (E$_{upt}$) does not display variations.}
\label{seds_breaks}
\end{figure*}

To investigate the spectral variability of 3C\,66A in detail, spectra have been obtained during
three intervals, shown in Fig.~\ref{lc_3c66a}. 
The division  takes into account the exposure for spectral analysis in  each interval and the range of different flux levels for the source. 
The assumes a monotonous change of spectral properties with the total flux even if several distinct flares are coadded. 
In the first interval there are only two observations performed in 2005, but the second one has an exposure of 52\,ks.
For each of the selected intervals, data are fitted with the \bk\ model with the Galactic absorption (\nh\ value taken from LAB survey), which  describe the overall spectrum well.
The spectral parameters are collected in Table~\ref{table_fits}.
The  plots presenting SEDs for all intervals are shown in Fig.~\ref{seds_breaks}.
In Sect.~\ref{catalogue} it was shown that  the average spectrum  is characterised by concave curvature and an  upturn energy of about 3.4\,keV. This value is an average one and the soft and hard intervals selected in Fig.~\ref{lc_3c66a} do not reflect precisely the synchrotron and the inverse Compton components for all times
The concave shape of the spectrum is clearly visible in the first interval. For the other intervals the upturn is not visible and the spectrum in the energy band  presents only a synchrotron component. Since the third epoch display the source at a fainter state compared to the first one, the absence of an upturn implies flux variability of the inverse Compton component. 

\begin{figure*}
\centering{\includegraphics[width=0.495\textwidth]{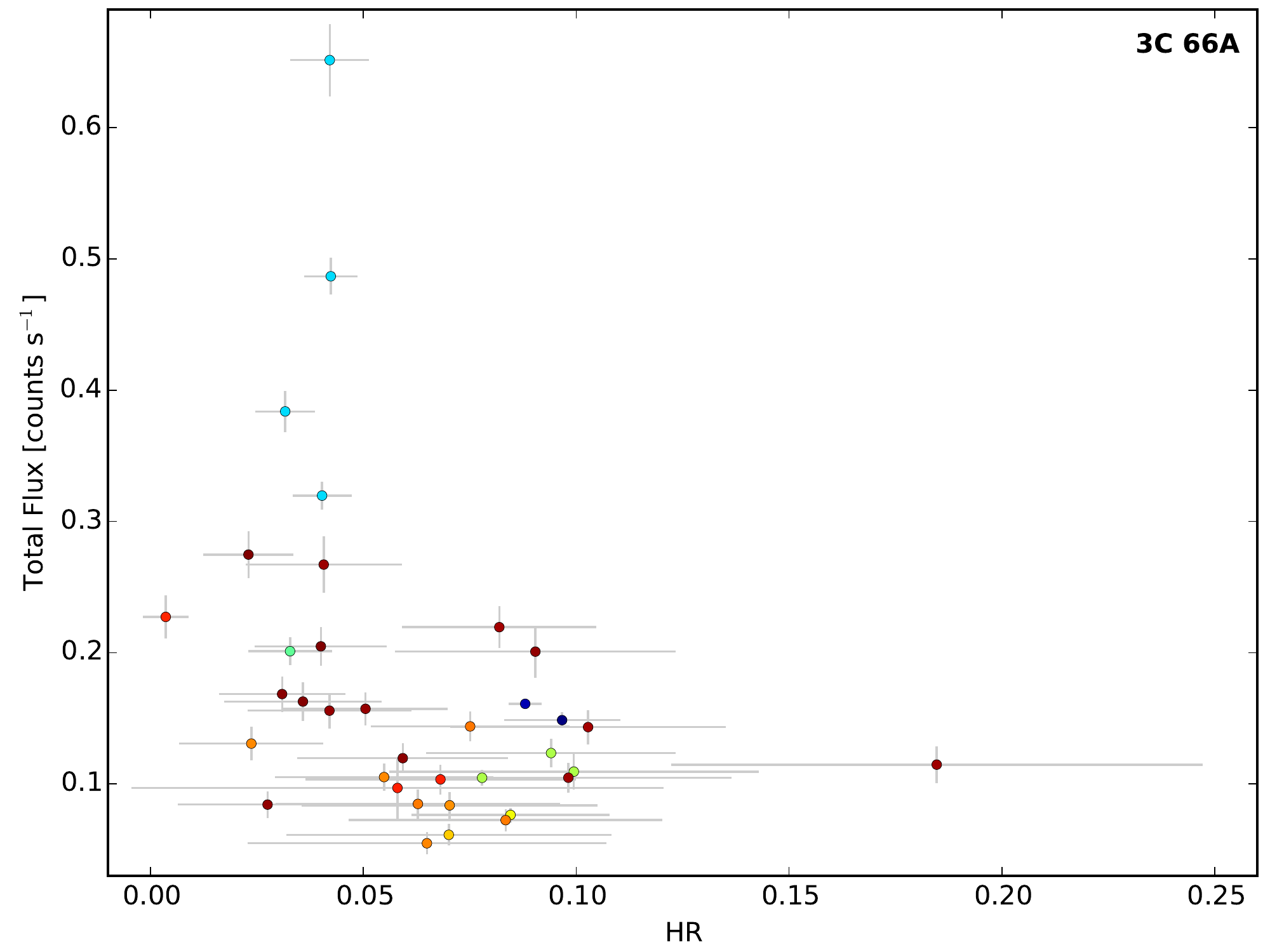}
\includegraphics[width=0.495\textwidth]{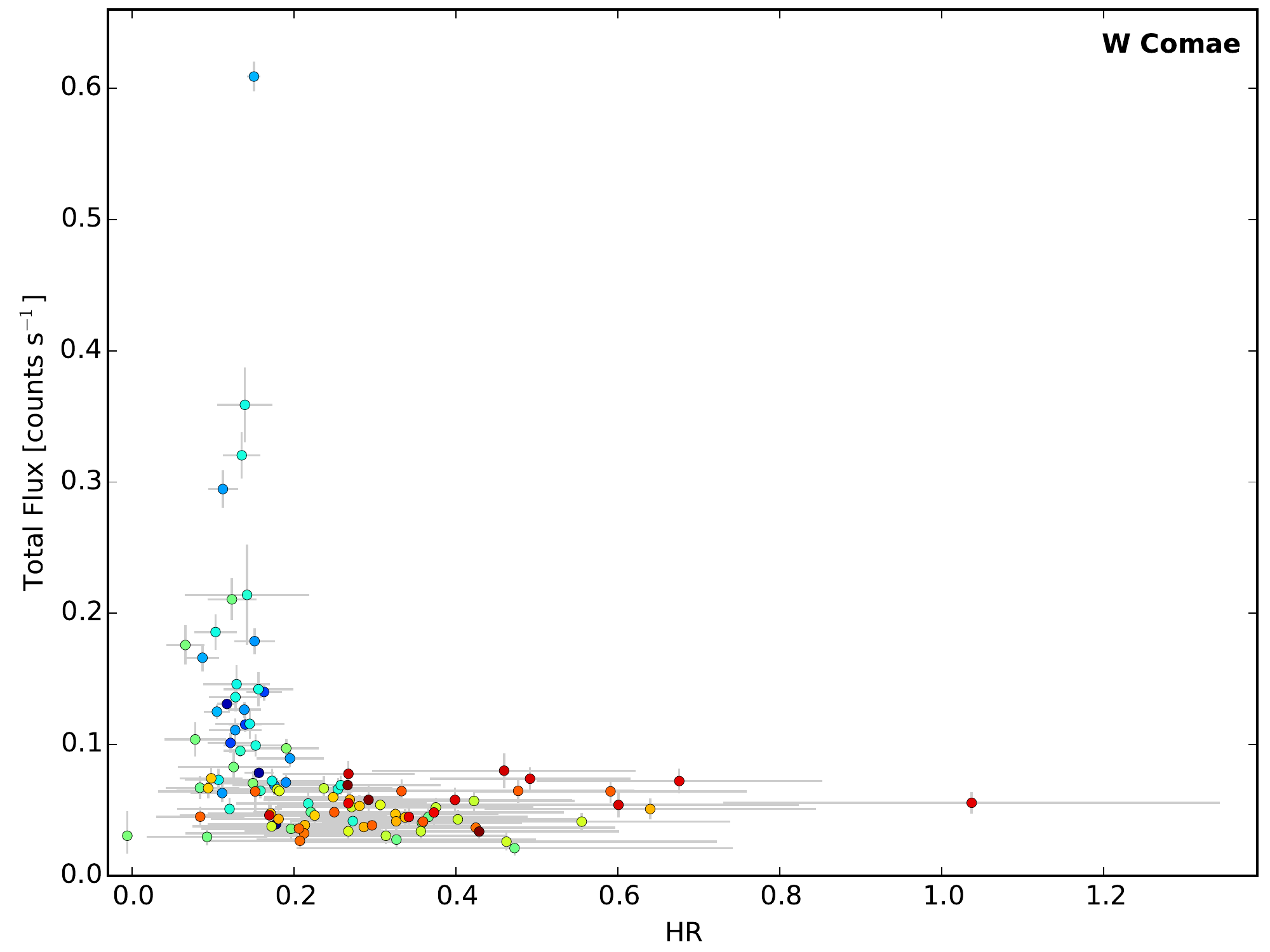}
\includegraphics[width=0.495\textwidth]{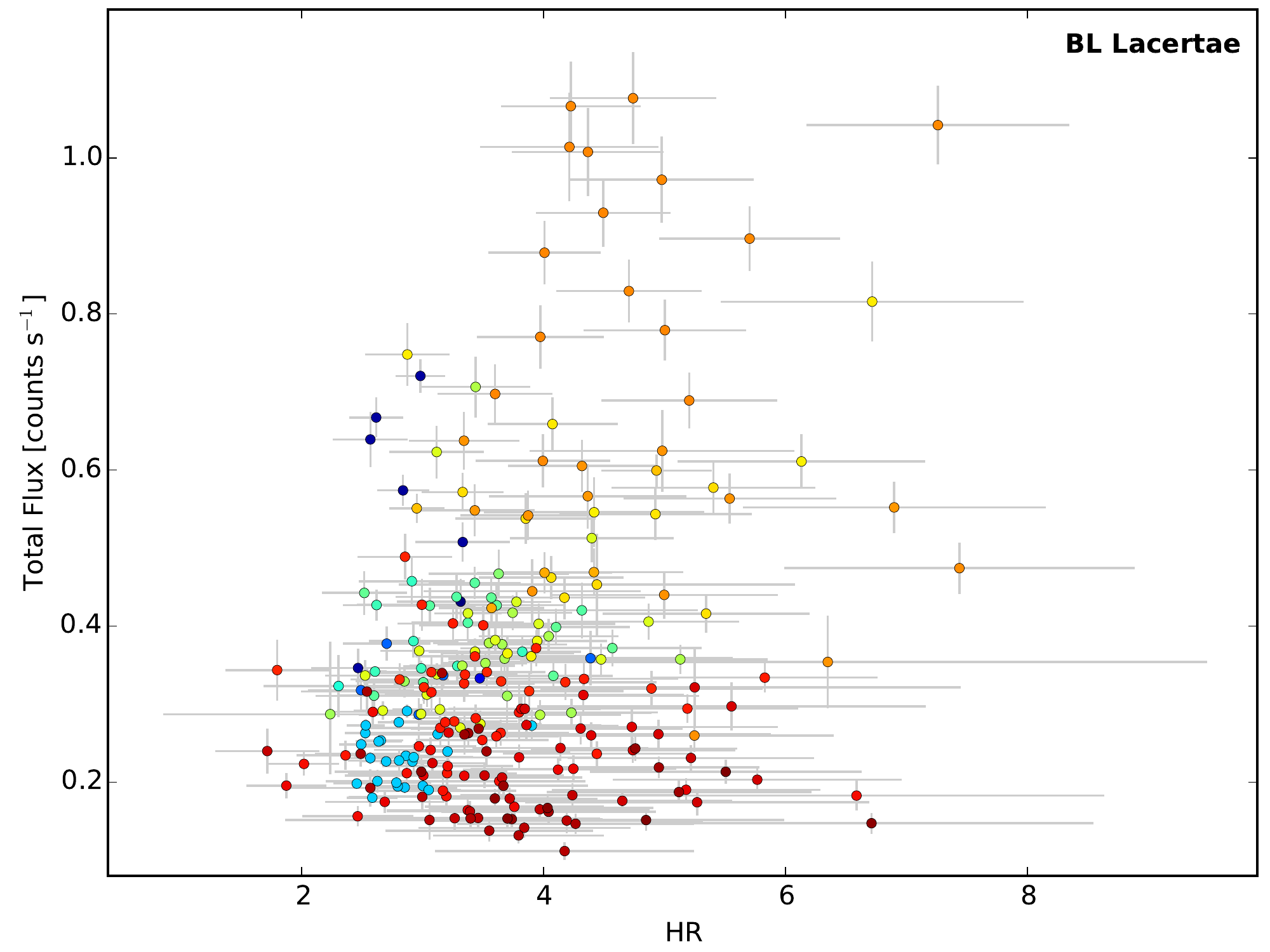}
\includegraphics[width=0.495\textwidth]{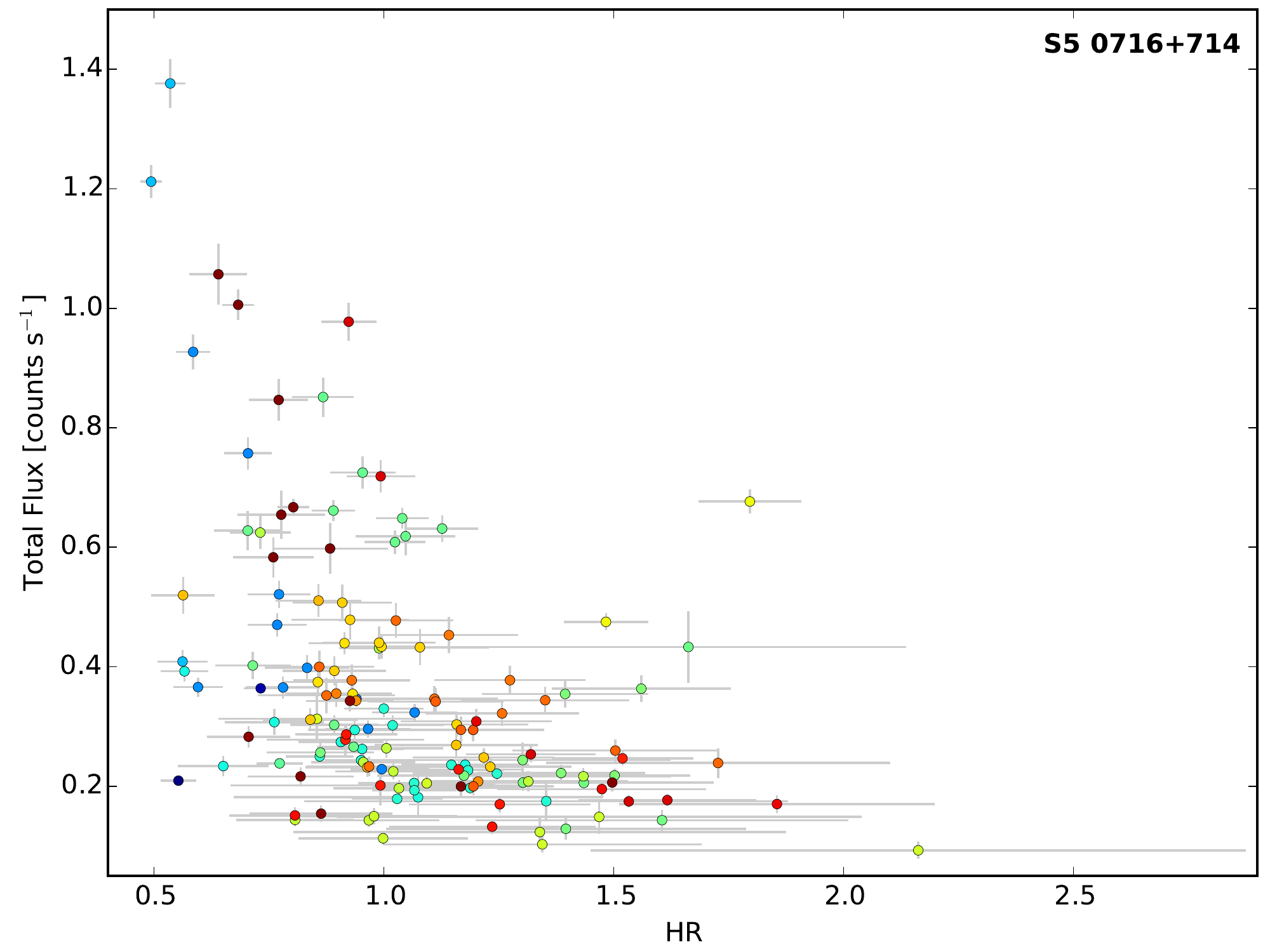}
\includegraphics[width=0.495\textwidth]{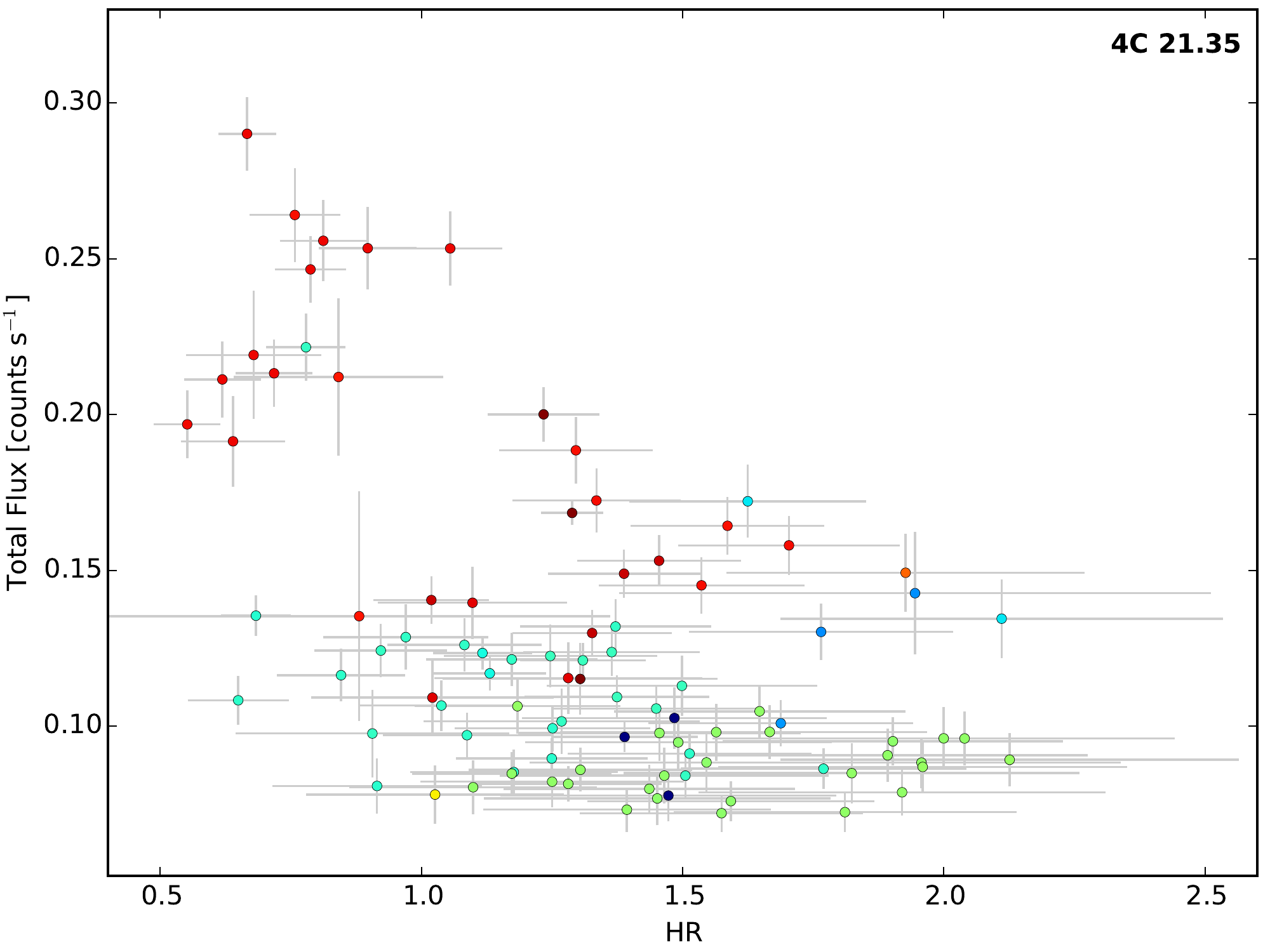}}
\caption{The comparison of the flux (total count rate) and the hardness ratio for 3C\,66A, W\,Comae, BL\,Lacertae, S5\,0716+714 and 4C\,21.35. The soft and hard bands are chosen as described in Sect.~\ref{variabilitysl} and presented in Fig.~\ref{lc_3c66a}-\ref{lc_4c2135}.
The colour of a given data point indicates the time of a given measurement. The earliest observations are denoted by dark blue symbols and the most recent ones depicted in red, with the rainbow colour scale normalized to the entire span of the \xrt\ observations of a given blazar. The same colour coding is used in Fig.\ref{lc_3c66a}-\ref{lc_4c2135}. }
\label{flux_hr}
\end{figure*}

Figure~\ref{flux_hr} presents the comparison of the hardness ratio and the total count rate. 
There is no clear trend between these quantities. The soft spectral state is observed in bright or faint, while
the hard state of the  source is for faint state.

\paragraph*{S5\,0716+714:}
The source, known for its variability, clearly manifests this feature in the longterm light curve presented in the upper panel of Fig.~\ref{lc_s50716}.
The middle panels of Fig.~\ref{lc_s50716}  shows the  hard and soft light curves, for which the energy bands are defined as 1.1-10\,keV and 0.3-1.1\,keV, respectively.
The longterm variability is  characterized by fractional variability amplitude for the total, soft and hard bands resulting in $F_\textit{var}=(61 \pm 1) \%$, $F_\textit{var}=(75 \pm 1) \%$ and $F_\textit{var}=(52 \pm 1) \%$, respectively. For the comparison the hardness ratio light curve (see bottom panel in Fig.~\ref{lc_s50716}) results in $F_\textit{var}=(22 \pm 2) \%$. Then, the most prominent variability is observed in the soft energy range.  

To study the spectral variability we divide the light curve into four intervals (see Fig.~\ref{lc_s50716}).
The first interval includes only two observations, but the first one has the time exposure of 19\,ks. 
Each of the spectrum, for selected intervals, can be well characterized with the \bk\ model with the Galactic absorption \nh\ value from DL survey. The spectral parameters are collected in Table~\ref{table_fits} and the corresponding plots are shown in Fig.~\ref{seds_breaks}. All the selected intervals have similar value of the break energy, between 0.9 and 1.4\,keV.
The overall the \bk\ fit to data results in the break energy of 1.1\,keV.
This indicates that division into the soft and the hard intervals with a good approximation reflects synchrotron and inverse Compton components.
We also investigate the spectral variability in the context of flux variation. Hardness ratio - total count rate comparison, presented in Fig.~\ref{flux_hr} indicates behaviour similar to the one observed in the case of 3C\,66A. 
It is also worth mentioning that in the case of this source we do not see clear harder-when-brighter trend or the opposite softer-when-brighter one. 
While the anticorrelation of the hardness ratio and the flux count rate in the case of S5\,0716+714 was previously reported in the X-ray regime by e.g. \cite{Foschini06, Ferrero06}.
The values of the spectral indices indicate significant spectral variability of the synchrotron component, which is not observed in the case of the inverse Compton part of the spectrum. 

\paragraph*{W~Comae:}
The observed count flux rate (see the upper panel of Fig.~\ref{lc_wcomea}) changes in the range of 0.1-0.7\,counts/s. The figure shows also the  soft (0.3-2.0\,keV) and the hard (2.0-10.0\,keV) energy band light curve as well as the hardness ratio evolution. 
The longterm variability of the blazar can be characterized with fractional variability amplitude resulting in $F_\textit{var}=(94 \pm 1) \%$, $F_\textit{var}=(69 \pm 2) \%$, $F_\textit{var}=(56 \pm 6) \%$, and $F_\textit{var}=(28 \pm 9) \%$  for the total, the soft, the hard and hardness ratio light curve respectively. These results show that the highest variability is observed in the energy range of 0.3-10\,keV.

As in the case of previously discussed sources, separate intervals are distinguished to investigate spectral features in the different states. In the case of W\,Comae two intervals are considered representing the low (the second interval) and the elevated (the first interval) flux level (see Fig.~\ref{lc_wcomea}). 
Data collected during both periods are fitted with the \bk\ model with the Galactic absorption \nh\ taken from the survey provided by \cite{Willingale13}. 
The spectral fit parameters are collected in Table~\ref{table_fits}, while all corresponding plots are shown in Fig.\ref{seds_breaks}. In both intervals the spectral shapes reveal a concave curvature. 
The selected intervals are characterized with different values of the break energy: 4.1\,keV and 1.2\,keV for the first and the second interval, respectively. It is worth reminding here that the crossing point in the \bk\ fit to all data is determined as 2.0\,keV. Because of this fact the soft and the  hard energy bands used for the light curves do not indicate clearly the synchrotron and inverse Compton components.
The values of the spectral indices reveal larger spectral variability in the soft energy band.
The comparison of the hardness ratio and the total count rate, presented in Fig.~\ref{flux_hr} exhibit similar behaviour as in the case of 3C\,66A and S5\,0716+714. 

\paragraph*{4C\,$+$21.35:}
The light curve presenting the longterm variability of the blazar (see the  upper panel of Fig.~\ref{lc_4c2135}) shows smaller changes in the observed count rate than in the case of other blazars discussed here. The variability observed in the energy range of 0.3-10\,keV can be described with fractional variability amplitude as high as $F_\textit{var}=(39 \pm 1) \%$.
The soft and the  hard energy bands, presented in the middle panels of Fig.~\ref{lc_4c2135}, are defined as: 0.3-1.2\,keV and 1.2-10\,keV.
The value of 1.2\,keV is the break energy point found in the average spectrum of this blazar.
To check whether the fractional variability amplitude depends on the energy band considered, this quantity has been also calculated for the soft and the hard energy bands resulting in: $F_\textit{var}=(56 \pm 1) \%$ and  $F_\textit{var}=(31 \pm 1) \%$, respectively. The bottom panel in Fig.~\ref{lc_4c2135}  show hardness ratio light curve, which is characterized by  $F_\textit{var}=(23 \pm 1) \%$. 

Three intervals have been distinguished for detailed spectral studies.
The spectral analysis has been repeated for each of the interval. Since in the case  of set of all observations data are well described with \bk\ model with the Galactic absorption, here also this model has been used. For this case \nh\ value taken from \cite{Willingale13} survey is used. Reduced $\chi^2$ values show that this model works well also for shorter intervals. 
All the spectral parameters of the \bk\ fit and the corresponding plots are collected in Table~\ref{table_fits} and shown in Fig.~\ref{seds_breaks}.
For the intervals selected the \bk\ fit results in the break energy of value 1.0-1.2\,keV. We can then conclude that the soft and the hard energy bands with a good approximation correspond to the synchrotron and the inverse Compton spectral components, respectively.
The comparison of the hardness ratio and the total count rate, presented in Fig.~\ref{flux_hr} shows complex behaviour, with individual substructures and indication for hysteresis effects visible within each.

\paragraph*{BL~Lacertae:}
The light curve of BL Lacertae (see upper panel of Fig.~\ref{lc_bllac}) in the energy range of 0.3-10\,keV shows significant variability exhibited in several outbursts. The longterm variability is characterized using fractional variability amplitude with the value of 
$F_\textit{var}=(52 \pm 1) \%$. As in the previous cases, we also consider the soft and the hard energy bands defined as 0.3-1.1\,keV and 1.1-10.0\,keV, respectively, as well as the hardness ratio.
The variability in the soft and the hard energy ranges is then given by $F_\textit{var}=(47 \pm 1) \%$ and $F_\textit{var}=(58 \pm 1) \%$, respectively. The variability amplitude is more pronounced in the higher energies. For the comparison the hardness ratio variability is quantified by $F_\textit{var}=(13 \pm 1) \%$.

To check if the crossing point, disentangling the synchrotron and the  inverse Compton components, is shifting for the  different flux states of the source, four intervals are distinguished (see the light curve in Fig.~\ref{lc_bllac}).
Each of the intervals represents different the X-ray behaviour of BL~Lacertae. Data in each interval are then fitted with the \bk\ model with the Galactic absorption \nh\ taken from the survey provided by \cite{Willingale13}. The fit parameters are collected in Table~\ref{table_fits} and the corresponding plots are shown in Fig.~\ref{seds_breaks}. Let us here remind that the \bk\ fit to all data results in the break energy of 1.1\,keV. 
For the shorter intervals the break energy is oscillating between 1.1-1.2\,keV. 
A clear upturn of the spectrum, disentangling two spectral components is visible in the case of the three intervals discussed. 

Furthermore, there is no significant shift of the crossing point to the higher energies with the increasing flux, which is observed in the other blazars discussed.
Hence, for this blazar the soft and hard energy bands in a good approximation correspond to the synchrotron and the inverse Compton components.  The synchrotron component is then characterized by larger spectral variability comparing to the inverse Compton one.
The hardness ratio seem to be correlated with the total count rate (see Fig.~\ref{flux_hr}), which is not surprising in the case of HBL type blazars, but it is not a common feature of the IBL ones.

\section{Summary and conclusions}\label{discussion}
In this paper we study spectra of 29 TeV~$\gamma$-ray emitting blazars including 22 HBL, 4 IBL type blazars, and 3 FSRQs observed with \xrt\ in the energy range of 0.3-10\,keV. 
The only one LBL blazar detected in  TeV $\gamma$-ray regime, (AP\,Librae) is not part of  the sample presented here.
Since only TeV blazars are included in the sample and HBL type objects are known to be the brightest in this regime, three groups of blazars included are not distributed uniformly. 

The spectra studied are averaged, which means that they include observations in the low and the high states, for data collected since the mission start up to October 2014. We focus only on the PC mode observations, which implicates that the highest outbursts are not included in the averaged spectra. 
This is the first spectral survey including 10\,years of observations of TeV emitting blazars collected with \xrt.

We conclude that the most of the spectra are well described using the curved \lp\ model. Only in a few cases we get better characteristics with the \po\ model. These are 1ES\,1101$-$232, Mrk\,180, and PKS\,2005$-$489. There are also a few cases of objects for which we got acceptable \po\ fit only if the free \nh\ value is used. These are mainly cases for which free value of the Galactic absorption is much higher than the ones provided in the  discussed surveys.

The curved \lp\ model has been used to describe the shape of radio-optical blazars spectra for the first time by \cite{Landau86}.
The first X-ray studies involving the \lp\ as the best model for spectral fits have been performed by \cite{Giommi02} who found this model as the preferable one for 50$\%$ of blazars studied. Also \cite{Massaro04a, Massaro04b, Tramacere07, Massaro08} successfully used this model in their X-ray spectral studies. 

The observed convex curvature in the X-ray spectra is likely to result from a single accelerated particle distribution
 \citep[e.g.][]{Massaro04b}.
The concave curvature observed may be caused by the fact that the X-ray range is a place, where synchrotron and inverse Compton components meet  and the curvature is a consequence of the spectral upturn. In the latter case the X-ray observations enable simultaneous study  of the low and the high  energy ends of the particle distribution function. 

\begin{figure*}
\centering{
\includegraphics[width=0.495\textwidth]{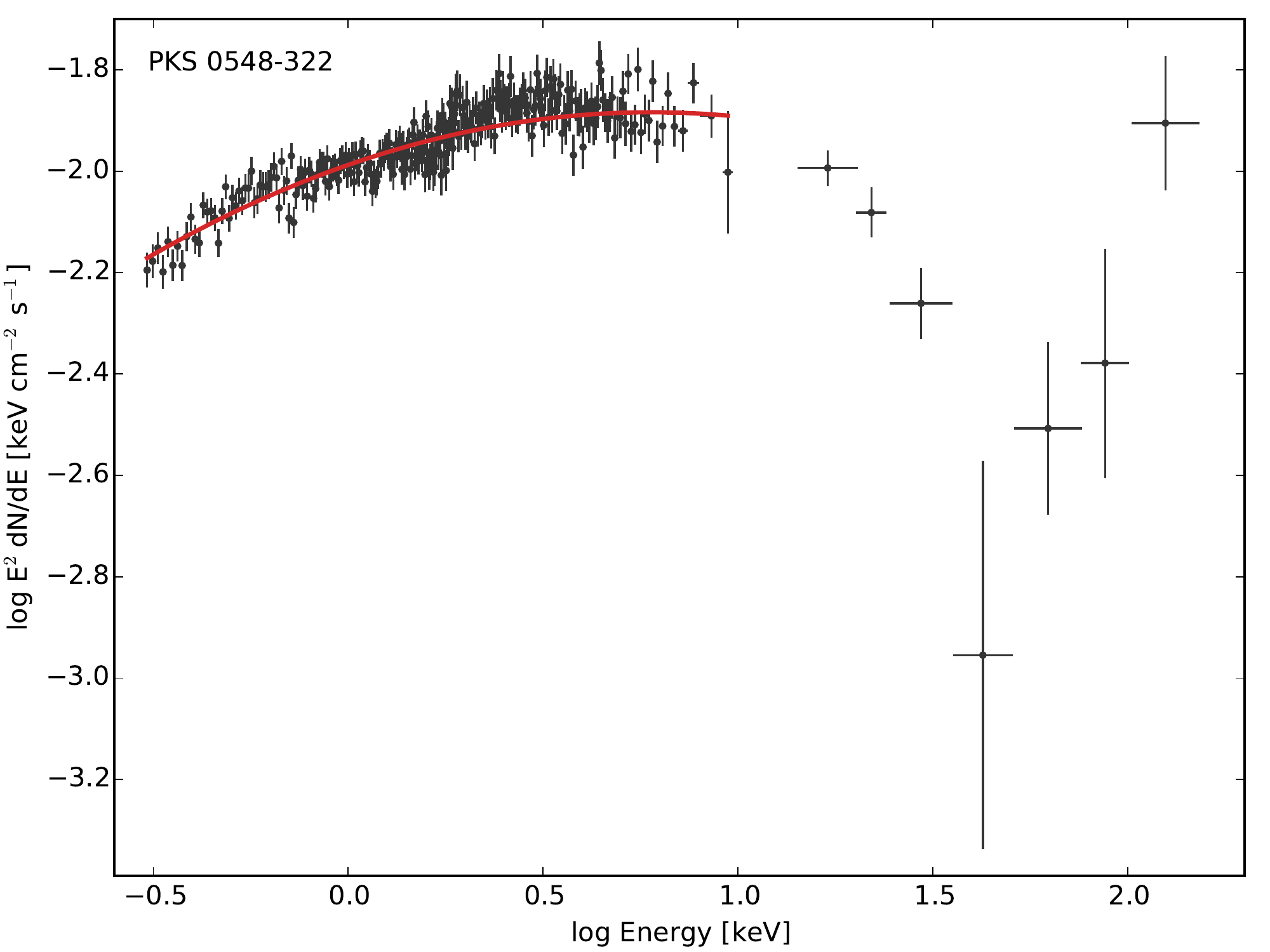}
\includegraphics[width=0.495\textwidth]{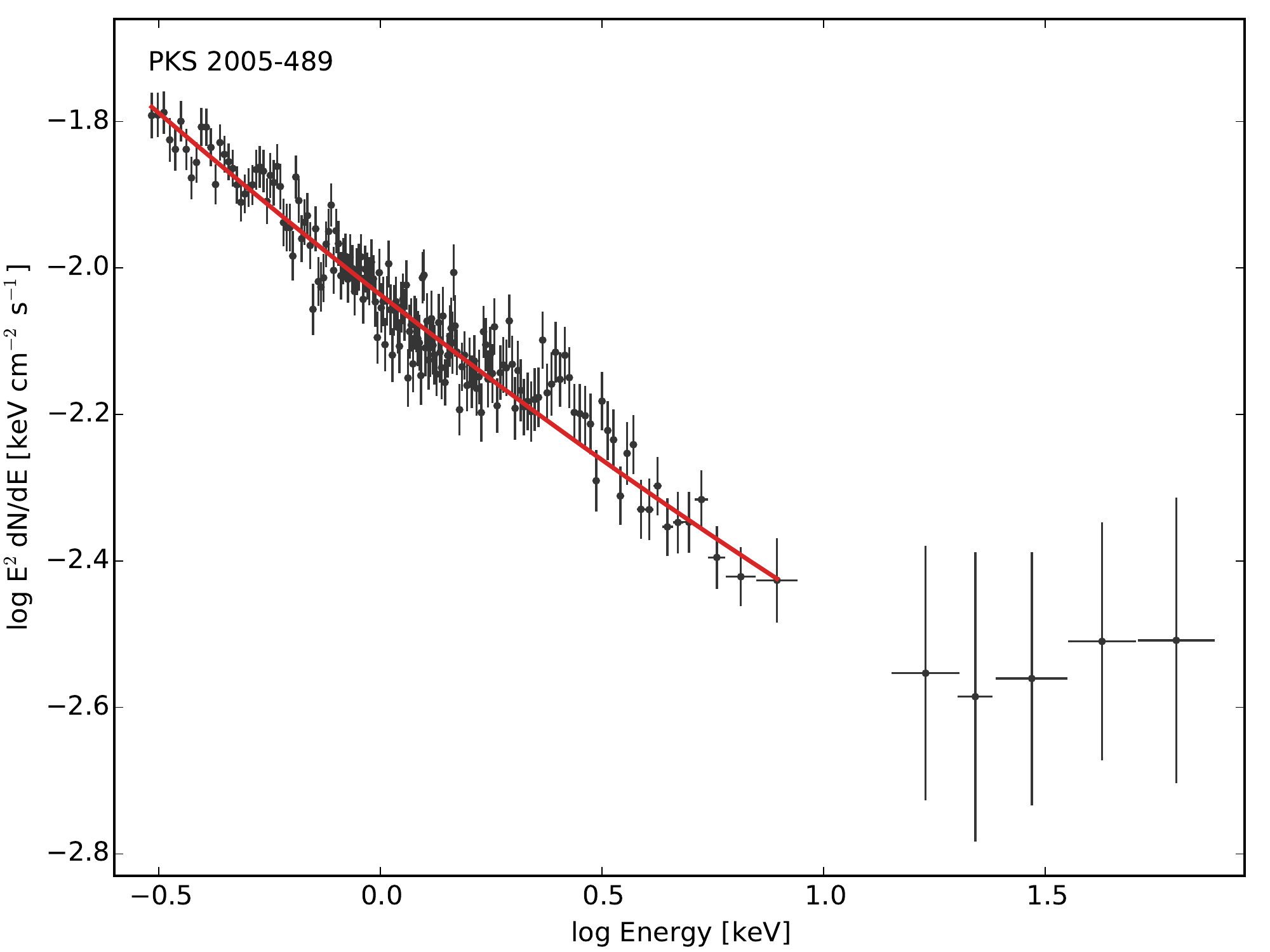}
}
\caption{The examples of joint spectral energy distributions including \xrt\ and \bat\ data for PKS 0548-322 and PKS\,2005-489. Each plot presents \xrt\ data  fitted with  a \lp\ model,  and the  \nh\ value taken from LAB survey, and BAT data from the 70-Month Hard X-ray Survey \citep{BAT_72m}. }
\label{xrt_bat}
\end{figure*}

Different values of \nh\ tested do not allow us to find one universal set. On the other hand we cannot conclude that any of mentioned \nh\ surveys can be used in each case to obtain good fit. But we can then summarize that each case should be treated separately, often including wider view on spectral features also in different wavelengths, see an example of two blazars presented in Fig.~\ref{xrt_bat}, where the \xrt\ observations are combined with the \bat\ ones.
Furthermore, our studies has shown that the Galactic column density value influence the overall spectral fit. In general larger \nh\ values result in softer spectral index, i.e.  larger value of $\Gamma$ or $\alpha$ parameter in the \po\ and the \lp\ model, respectively.
Also in  most cases for the \lp\ model fitted the curvature parameter is larger for smaller \nh\ values. 
Hence, we can summarize that the choice has significant influence on the X-ray spectral features. It is also worth  mentioning that \ch\ is not a good indicator, which could be used to distinguish the most accurate value of the Galactic column density. Let us here remind the example of  1ES\,0033+595, for which very different values of \nh\  result in the significantly different spectral fit parameters and very similar values of \ch\ (see Fig.~\ref{seds_lp} and \ref{seds_po} (in the online material), and Table~\ref{table_fitpar_lp} and \ref{table_fitpar_po} (in the online material) for plots and spectral fit parameters for the \lp\ and the \po\ models). In the mentioned case too small value of \nh\ e.g. the one taken from LAB survey result in the spectral shape with the artificial bump which cannot be easily explained.
Very similar situation take place for the cases of blazars such like 1ES\,0502+675, 1ES\,0647+250 or 1ES\,1959+650.

The \lp\ model is described by the $\alpha$ parameter, which corresponds to the spectral index and $\beta$ one, which characterize the spectral curvature. We found the spectral indices $\alpha$ in the range of  1.4-2.6.

The performed spectral studies of 29 $\gamma$-ray emitting blazars allow us to distinguish only 5 blazars for which the concave spectral shape indicates that in the X-ray regime synchrotron and inverse Compton components meet. 
Since the spectral upturn is expected in all blazar SEDs and we focused on the energy range of one magnitude,  we then conclude that the spectral upturns are distributed in the broad range of energies
The second part of the paper includes detailed  temporal and spectral studies  focused on 5 sources for which concave shape of the spectrum indicates that  the synchrotron and the inverse Compton components meet in the X-ray regime.
The mentioned sources are: 4C\,$+$21.35, S5\,0716+714, BL\,Lacertae, 3C\,66A, and W~Comae.
For each  of the objects  the preferable model to describe the spectral shape is  the \lp\ with a concave curvature.  
For these five blazars with the upturn, we fit spectra with the \bk\ model, which helps us to find the crossing point for the average spectrum, disentangling the synchrotron and the inverse Compton components.
For all sources the upturn point is clearly visible in the averaged spectrum for the dataset including all data analysed. 
But for the case of shorter time intervals synchrotron and inverse Compton components are not always visible.
For example 3C\,66A for A and B intervals two components are clearly visible, while for the case of the third distinguished interval, the spectral shape reveals only the synchrotron part of the spectrum.
It is also worth  underlining that the upward curvature was previously reported for S5\,0716+714 \citep{Cappi94,Giommi99,Tagliaferri03,Donato05,Ferrero06}, BL\,Lacertae \citep{Tanihata00,Ravasio02,Donato05}, 3C\,66A \citep{Donato05}, and W~Comae \citep{Tagliaferri00, Donato05}, 
while this is the first case when the upturn between the synchrotron and the inverse Compton component is found for 4C\,$+$21.35.
Using \xrt\ observations of blazars we found five objects with the spectral upturn. Also, there are objects for which single \xrt\ observations do not reveal this feature, but joint spectrum, that include \xrt\ and \bat\ data show such an upturn. But since \bat\ spectra are longterm-averaged it is difficult to quantify the number of such objects. As an example Fig.~\ref{xrt_bat} show joint fits for PKS\,0548-322 and PKS\,2005-489.

The longterm variability of those five blazars has been characterized with fractional variability amplitude. This quantity is also calculated for the distinguished the  soft and the  hard energy bands as well as for the hardness ratio evolution. 
The soft and the hard energy band corresponds to the synchrotron and the  inverse Compton components in the average spectral fit. 
For 4C\,$+$21.35, S5\,0716+714, 3C\,66A, and W~Comae the largest value of $F_\textit{var}$ is obtained for the soft energy range, while the lowest variability is observed for the hardness ratio light curve. 
Only for BL\,Lacertae the largest value of $F_{var}$ is for the hard energy range. 

We also investigate the spectral variability by dividing the light curves into shorter intervals and performing spectral analysis for smaller data sets. In this case only the \bk\ model with the Galactic absorption value taken from LAB survey is tested. 
In the case of S5\,0716+714, 4C\,$+$21.35 and BL\,Lacertae the energy break value are consistent within errors with the value obtained for the average spectrum. Hence, we conclude that the distinguished soft and the hard energy bands corresponds to the  synchrotron and the inverse Compton components, respectively.
In the case of 3C\,66A, and W~Comae the break energy values for the short intervals and the average spectrum are different, so the soft and the hard bands can not be unambiguously associated with the synchrotron and the inverse Compton components.
Furthermore the comparison of the spectral indices for the \bk\ fit for the shorter intervals revealed significant variability in the soft energy range, while in the hard one the variability is not noticed in the case of 3C\,66A.

We also compared the hardness ratio, calculated as a ratio of count rates in the  hard and the  soft energy bands, with the total intensity. 
This comparison revealed different behaviours of in  HR-total intensity diagrams.
For three of studied blazars, namely 3C\,66A, W\,Comae and S5\,0716+714 for the soft spectrum (smaller values of HR) total intensity of the sources changes in the whole range of brightness.
While when the source is faint, different values of HR can be observed. 
Such a behaviour suggests that in the faint state of the source the influence of the  inverse Compton component is more significant.
For BL\,Lacertae HR-total intensity diagram exhibit harder-when-brighter trend. 
Such a behaviour is a common feature for HBL type blazars \citep[e.g.][]{Pian98, Zhang2005, Zhang06b, Abramowski2012_2155}, while IBL and LBL ones often exhibits softer-when-brighter pattern \citep[e.g.][]{Giommi99, Ravasio02, Ferrero06}. 
The case of 4C\,21.35 is the most complex one with individual substructures and an indication for hysteresis effects visible within each. Such a hysteresis trend visible in HR-flux diagram can be related to the  changes of the  acceleration and cooling time scales \citep{Kirk98}.

\begin{figure*}
\centering{\includegraphics[width=0.92\textwidth]{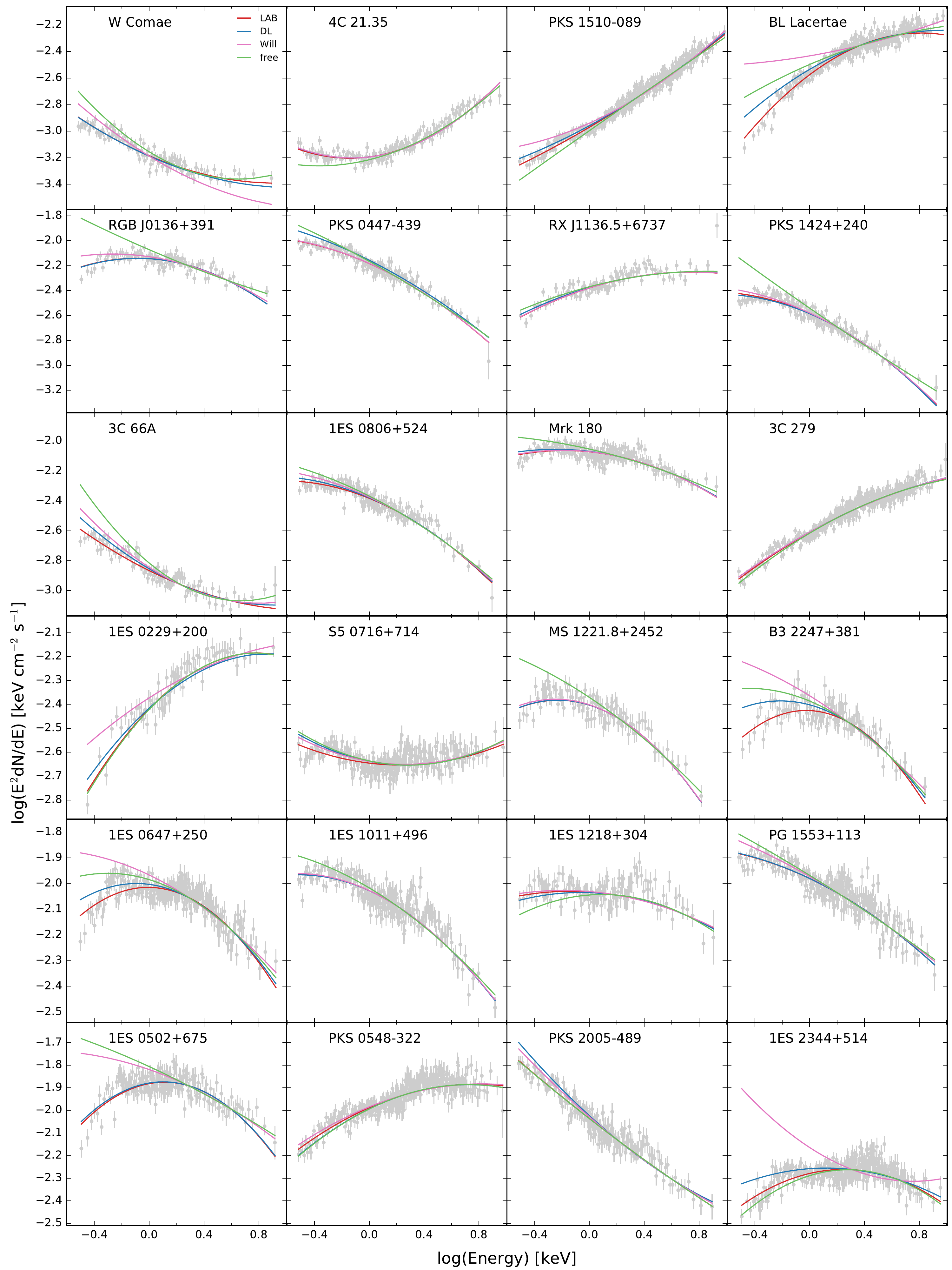}}
\caption{The X-ray spectral energy distributions of 29 selected blazars sorted by increasing flux density. The names of the sources are given in the upper part of each panel. 
Four  \lp\ spectral models with different values of \nh\ are presented   in different colours: red --  \nh\  taken from the LAB survey
\citep{Kalberla05},
blue --  \nh\ value taken from 
\citet{Dickey90}, 
magenta --  \nh\ value taken from 
\citet{Willingale13} 
and green --   free \nh\ value. 
Data (grey points) presented in the figure are those  fitted with the \lp\ model  with the Galactic absorption value taken from LAB survey.}
\label{seds_lp}
\end{figure*}

\begin{figure*}
\centering{\includegraphics[width=0.92\textwidth]{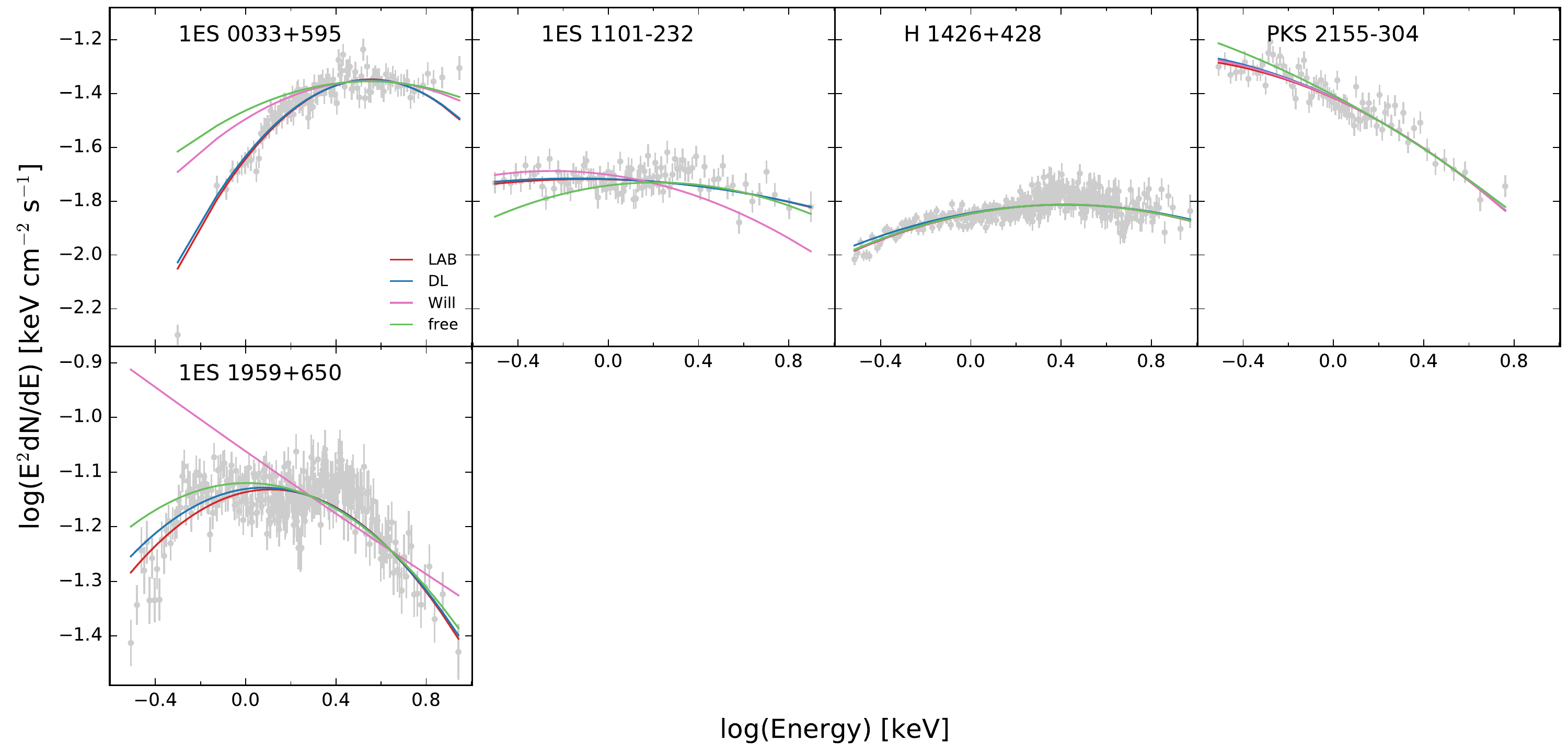}}
\contcaption{}
\label{seds_lp}
\end{figure*}

\begin{figure*}
\caption{The X-ray spectral energy distributions of 29 selected blazars sorted by increasing flux density. The names of the sources are given in the upper part of each panel. 
Four \po\ models using different values of \nh\ are presented   using different colours: red --  \nh\  taken from the LAB survey
\citep{Kalberla05}, blue --  \nh\ value taken from \citet{Dickey90}, magenta --  \nh\ value taken from \citet{Willingale13} and green --   free \nh\ value. 
Data (grey points) presented in the figure are those  fitted with the \po\ model  with the Galactic absorption value taken from LAB survey. The figure is available in the online material only.}
\label{seds_po}
\end{figure*}

\begin{figure*}
\caption{The X-ray spectra fitted with the \lp\ model with the Galactic absorption value taken from LAB survey and the ratios (data points divided by the folded model). The following plots present spectra for: PKS\,2005-489, PKS\,2155-304. BL\,Lacertae, B3\,2247+381, 1ES\,2344+514. The name of each object is also given in the upper right corner of each plot. }
\label{spec_5}
\end{figure*}

\clearpage
\newpage


\begin{landscape}
\begin{table}

\caption{The summary of TeV blazars.  (1) The object name, (2) the right
ascension, (3) the declination, (4) the type of the blazar,  (5) the redshift,
(6) the integrated flux for \xrt\ observations in PC mode in the energy range
of 0.3-10\,keV,  (7) the exposure for \xrt\ observations in PC mode in the
energy range of 0.3-10\,keV,  (8) the product of (6) and (7),  (9) the
information if the blazar has been selected for the further detailed studies;
selected objects are marked with + sign.}
\label{table_tevblazars}

\centering

\begin{tabular}{rcccccccc}

\hline
Object & RA & DEC& type  &  redshift &   PC flux   & PC exposure  & PC flux $\cdot$ exposure & notes   \\
        &  $^h$\,$^m$\,$^s$  &  $^\circ$\,' \,'' &    &          &   10$^{-12}$\,erg\,cm$^{-2}$\,s$^{-1}$     &  ks   &  10$^{-11}$\,erg\,cm$^{-2}$ &      \\

   (1)& (2) & (3) & (4) & (5) & (6) & (7) & (8) & (9)     \\ 
\hline

SHBL\,J001355.9$-$185406\tabref{1} & 00 13 52.0 & $-$18 53 29 & HBL  & 0.095\tabref{a}   & 12.30  & 14.4  & 1.77  & \\
KUV\,00311$-$1938\tabref{2}        & 00 33 34.2 & $-$19 21 33 & HBL  & 0.61\tabref{b}    & 8.10   & 19.3  & 1.56  & \\
1ES\,0033$+$595\tabref{3}          & 00 35 52.6 & $+$59 50 05 & HBL  & $>$0.24\tabref{c} & 109.90 & 15.2  & 16.70 & $+$ \\
RGB\,J0136$+$391\tabref{4}         & 01 36 32.5 & $+$39 06 00 & HBL  & Unknown                     & 23.80  & 18.8  & 4.47  & $+$ \\
RGB\,J0152$+$017\tabref{5}         & 01 52 33.5 & $+$01 46 40 & HBL  & 0.08\tabref{d}    & 6.58   & 27.1  & 1.78  & \\
3C\,66A\tabref{6}                  & 02 22 41.6 & $+$43 02 35 & IBL  & 0.41\tabref{e}    & 5.14   & 92.4  & 4.75  & $+$ \\
1ES\,0229$+$200\tabref{7}          & 02 32 53.2 & $+$20 16 21 & HBL  & 0.14\tabref{f}    & 20.70  & 22.4  & 4.64  & $+$ \\
PKS\,0301$-$243\tabref{8}          & 03 03 23.5 & $-$24 07 36 & HBL  & 0.2657\tabref{c}  & 8.38   & 25.2  & 2.11  & \\
IC\,310\tabref{9}                  & 03 16 43.0 & $+$41 19 29 & HBL  & 0.0189\tabref{g}  & 6.60   & 68.0  & 4.49  & \\
RBS\,0413\tabref{10}               & 03 19 47.0 & $+$18 45 42 & HBL  & 0.19\tabref{h}    & 10.70  & 7.3   & 0.78  & \\
1ES\,0347$-$121\tabref{11}         & 03 49 23.0 & $-$11 58 38 & HBL  & 0.188\tabref{c}   & 43.70  & 3.2   & 1.40  & \\
1ES\,0414$+$009\tabref{12}         & 04 16 53.0 & $+$01 05 20 & HBL  & 0.287\tabref{i}   & 18.30  & 6.2   & 1.13  & \\
PKS\,0447$-$439\tabref{13}         & 04 49 28.2 & $-$43 50 12 & HBL  & 0.107\tabref{j}   & 24.60  & 50.5  & 12.42 & $+$ \\
1ES\,0502$+$675\tabref{14}         & 05 07 56.2 & $+$67 37 24 & HBL  & 0.341\tabref{k}   & 49.60  & 39.9  & 19.79 & $+$ \\
VER\,J0521$+$211\tabref{15}        & 05 21 45.0 & $+$21 12 51 & IBL  & 0.108\tabref{l}   & \multicolumn{3}{|c|}{Not enough data in the PC mode  to constrain spectral parameters.} & \\
PKS\,0548$-$322\tabref{16}         & 05 50 38.4 & $-$32 16 12 & HBL  & 0.069\tabref{m}   & 55.70  & 103.1 & 57.43 & $+$ \\
RX\,J0648.7$+$1516\tabref{17}      & 06 48 45.6 & $+$15 16 12 & HBL  & 0.179\tabref{n}   & 12.40  & 10.6  & 1.31  & \\
1ES\,0647$+$250\tabref{18}         & 06 50 46.5 & $+$25 03 00 & HBL  & 0.45\tabref{o}    & 31.80  & 71.4  & 22.71 & $+$ \\
RGB\,J0710$+$591\tabref{19}        & 07 10 26.4 & $+$59 09 00 & HBL  & 0.125\tabref{n}   & \multicolumn{3}{|c|}{Not enough data  in the PC mode to constrain spectral parameters.} & \\
S5\,0716$+$714\tabref{20}          & 07 21 53.4 & $+$71 20 36 & IBL  & 0.31\tabref{h}    & 11.18  & 231.0 & 25.83 & $+$ \\
1ES\,0806$+$524\tabref{21}         & 08 09 59.0 & $+$52 19 00 & HBL  & 0.137\tabref{p}   & 14.68  & 51.9  & 7.62  & $+$ \\
RBS\,0723\tabref{22}               & 08 47 12.9 & $+$11 33 50 & HBL  & 0.198\tabref{p}   & 19.60  & 7.1   & 1.39  & \\
1RXS\,J101015.9$-$311909\tabref{23}& 10 10 15.0 & $-$31 18 18 & HBL  & 0.143\tabref{q}   & 13.0   & 3.7   & 0.48  & \\
S4\,0954$+$65\tabref{24}           & 09 58 47.0 & $+$65 33 55 & FSRQ & 0.368\tabref{r}   & 3.48   & 68.8  & 2.39  & \\
1ES\,1011$+$496\tabref{25}         & 10 15 04.1 & $+$49 26 01 & HBL  & 0.212\tabref{s}   & 34.30  & 58.5  & 20.07 & $+$ \\
1ES\,1101$-$232\tabref{26}         & 11 03 36.5 & $-$23 29 45 & HBL  & 0.186\tabref{h}   & 90.10  & 12.5  & 11.26 & $+$ \\
Markarian\,421\tabref{27}          & 11 04 19.0 & $+$38 11 41 & HBL  & 0.031\tabref{t}   & \multicolumn{3}{|c|}{Not enough data in the PC mode to constrain spectral parameters.} & \\
Markarian\,180\tabref{28}          & 11 36 26.4 & $+$70 09 27 & HBL  & 0.046\tabref{r}   & 37.90  & 50.7  & 19.22 & $+$ \\
RX\,J1136.5$+$6737\tabref{29}      & 11 36 30.1 & $+$67 37 04 & HBL  & 0.1342\tabref{p}  & 20.70  & 21.0  & 4.35  & $+$ \\
1ES\,1215$+$303\tabref{30}         & 12 17 48.5 & $+$30 06 06 & HBL  & 0.237\tabref{u}   & 6.11   & 47.5  & 2.90  & \\
1ES\,1218$+$304\tabref{31}         & 12 21 26.3 & $+$30 11 29 & HBL  & 0.183\tabref{p}   & 33.30  & 21.6  & 7.19  & $+$ \\
W\,Comae\tabref{32}                & 12 21 31.7 & $+$28 13 59 & IBL  & 0.103\tabref{p}   & 3.13   & 166.3 & 5.21  & $+$ \\
MS\,1221.8$+$2452\tabref{33}       & 12 24 24.2 & $+$24 36 24 & HBL  & 0.218\tabref{v}   & 15.60  & 28.2  & 4.40  & $+$ \\
4C\,$+$21.35\tabref{34}            & 12 24 54.4 & $+$21 22 46 & FSRQ & 0.435\tabref{m}   & 4.95   & 174.3 & 8.63  & $+$ \\
S3\,1227$+$25\tabref{35}           & 12 30 14.1 & $+$25 18 07 & IBL  & 0.135\tabref{w}   & \multicolumn{3}{|c|}{Not enough data in the PC mode to constrain spectral parameters.} & \\
3C\,279\tabref{36}                 & 12 56 11.1 & $-$05 47 22 & FSRQ & 0.538\tabref{x}   & 15.03  & 595.6 & 89.52 & $+$ \\
1ES\,1312$-$423\tabref{37}         & 13 14 58.5 & $-$42 35 49 & HBL  & 0.105\tabref{h}   & 28.20  & 4.8   & 1.35  & \\
PKS\,1424$+$240\tabref{38}         & 14 27 00.0 & $+$23 47 40 & HBL  & 0.390\tabref{y}   & 9.03   & 49.8  & 4.50  & $+$ \\
H\,1426$+$428\tabref{39}           & 14 28 32.6 & $+$42 40 21 & HBL  & 0.129\tabref{p}   & 59.20  & 123.6 & 73.17 & $+$ \\
1ES\,1440$+$122\tabref{40}         & 14 42 48.3 & $+$12 00 40 & HBL  & 0.1631\tabref{p}  & 19.20  & 7.6   & 1.46  & \\
PKS\,1441$+$25\tabref{41}          & 14 43 56.9 & $+$25 01 44 & FSRQ & 0.939\tabref{z}   & 0.55   & 3.7   & 0.02  & \\
\hline
\end{tabular}
\end{table}
\end{landscape}

\clearpage
\newpage

\begin{landscape}
\begin{table}

\contcaption{}

\begin{centering}

\begin{tabular}{rcccccccc}

\hline
   (1)& (2) & (3) & (4) & (5) & (6) & (7) & (8) & (9)     \\ 
\hline

PKS\,1510$-$089\tabref{42}         & 15 12 52.2 & $-$09 06 21 & FSRQ & 0.360\tabref{m}   & 10.51  & 362.0 & 38.05 & $+$ \\
AP\,Librae\tabref{43}              & 15 17 41.8 & $-$24 22 19 & LBL  & 0.049\tabref{c}   & 6.36   & 32.7  & 2.08  & \\
PG\,1553$+$113\tabref{44}          & 15 55 44.7 & $+$11 11 41 & HBL  & $>$0.62\tabref{aa}& 43.70  & 86.5  & 37.80 & $+$ \\
Markarian\,501\tabref{45}          & 16 53 52.2 & $+$39 45 37 & HBL  & 0.037\tabref{x}   & \multicolumn{3}{|c|}{Not enough data in the PC mode to constrain spectral parameters.} & \\
Markarian\,501\tabref{45}          & 16 53 52.2 & $+$39 45 37 & HBL  & 0.037\tabref{x}   & 0.0    & 0.0   & 0.0   & \\
H\,1722$+$119\tabref{46}           & 17 25 04.3 & $+$11 52 15 & HBL  & 0.018\tabref{h}   & 10.65  & 2.7   & 0.29  & \\
1ES\,1727$+$502\tabref{47}         & 17 28 18.6 & $+$50 13 10 & HBL  & 0.055\tabref{ab}  & 16.30  & 5.5   & 0.90  & \\
1ES\,1741$+$196\tabref{48}         & 17 43 57.8 & $+$19 35 09 & HBL  & 0.083\tabref{ac}  & 11.70  & 18.8  & 2.20  & \\
HESS\,J1943$+$213\tabref{49}       & 19 43 55.0 & $+$21 18 08 & HBL  & $>$0.14\tabref{ad}& 20.60  & 13.7  & 2.82  & \\
1ES\,1959$+$650\tabref{50}         & 19 59 59.8 & $+$65 08 55 & HBL  & 0.047\tabref{x}   & 270.00 & 30.2  & 81.54 & $+$ \\
MAGIC\,J2001$+$435\tabref{51}      & 20 01 13.5 & $+$43 53 02 & IBL  & $<$ 0.2\tabref{ae}& 2.11   & 62.2  & 1.31  & \\
PKS\,2005$-$489\tabref{52}         & 20 09 27.0 & $-$48 49 52 & HBL  & 0.071\tabref{m}   & 31.00  & 50.1  & 15.53 & $+$ \\
PKS\,2155$-$304\tabref{53}         & 21 58 52.7 & $-$30 13 18 & HBL  & 0.116\tabref{af}  & 97.10  & 89.8  & 87.20 & $+$ \\
BL\,Lacertae\tabref{54}            & 22 02 43.3 & $+$42 16 40 & IBL  & 0.069\tabref{x}   & 14.15  & 365.6 & 51.73 & $+$ \\
B3\,2247$+$381\tabref{55}          & 22 50 06.6 & $+$38 25 58 & HBL  & 0.119\tabref{ac}  & 10.60  & 42.3  & 4.48  & $+$ \\
RGB\,J2243$+$203\tabref{56}        & 22 43 54.7 & $+$20 21 04 & HBL  & Unknown                     & 4.94   & 9.5   & 0.47  & \\
1ES\,2344$+$514\tabref{57}         & 23 47 04.9 & $+$51 42 17 & HBL  & 0.044\tabref{t}   & 20.38  & 143.9 & 29.33 & $+$ \\
H\,2356$-$309\tabref{58}           & 23 59 09.4 & $-$30 37 22 & HBL  & 0.165\tabref{ag}  & 5.68   & 17.1  & 0.97  & \\

\hline
\end{tabular}

\end{centering}

TeV detection references:\hfill\break
\tabref{1}\cite{Abramowski13_0011355},
\tabref{2}\cite{Becherini12_00311},
\tabref{3}\cite{Aleksic15_0033},
\tabref{4}\cite{Mazin2012_0136},
\tabref{5}\cite{Aharonian08_0152},
\tabref{6}\cite{Aliu09_66a},
\tabref{7}\cite{Aharonian07_0229},
\tabref{8}\cite{Aharonian13_0301},
\tabref{9}\cite{Aleksic10_310},
\tabref{10}\cite{Aliu12_0413},
\tabref{11}\cite{Aharonian07_0347},
\tabref{12}\cite{Abramowski12_0414},
\tabref{13}\cite{Abramowski13_0447},
\tabref{14}\cite{Benbow11_0502},
\tabref{15}\cite{Archambault13_0521},
\tabref{16}\cite{Aharonian10_0548},
\tabref{17}\cite{Aliu11_0648},
\tabref{18}\cite{Lotto12_0647},
\tabref{19}\cite{Acciari10_0710},
\tabref{20}\cite{Anderhub09_0716},
\tabref{21}\cite{Acciari09_0806},
\tabref{22}\cite{Mirzoyan14_0723},
\tabref{23}\cite{Abramowski12_101015},
\tabref{24}\cite{Mirzoyan_s40954},
\tabref{25}\cite{Albert07_1011},
\tabref{26}\cite{Aharonian07_1101},
\tabref{27}\cite{Punch92_421},
\tabref{28}\cite{Albert06_180},
\tabref{29}\cite{Mirzoyan14_j1136},
\tabref{30}\cite{Aleksic12_es1215},
\tabref{31}\cite{Albert06_1218},
\tabref{32}\cite{Acciari08_wcom},
\tabref{33}\cite{Cortina13_1221},
\tabref{34}\cite{Aleksic12_1222},
\tabref{35}\cite{Mukherjee1227},
\tabref{36}\cite{Errando08_279},
\tabref{37}\cite{Abramowski13_1312},
\tabref{38}\cite{Acciari10_1424},
\tabref{39}\cite{Horan02_1426},
\tabref{40}\cite{Ong10_1440},
\tabref{41}\cite{Mukherjee1441},
\tabref{42}\cite{Abramowski13_1510},
\tabref{43}\cite{Abramowski14_aplib},
\tabref{44}\cite{Aharonian06_1553},
\tabref{45}\cite{Quinn96_501},
\tabref{46}\cite{Cortina13_1722},
\tabref{47}\cite{Aleksic14_1727},
\tabref{48}\cite{Berger11_1741},
\tabref{49}\cite{Abramowski11_1943},
\tabref{50}\cite{Nishiyama99_1959},
\tabref{51}\cite{Mariotti10_2001},
\tabref{52}\cite{Aharonian05_2005},
\tabref{53}\cite{Chadwick99_2155},
\tabref{54}\cite{Albert07_bllac},
\tabref{55}\cite{Aleksic12_2247},
\tabref{56}\cite{Holder2243},
\tabref{57}\cite{Catanese98_2344},
\tabref{58}\cite{Aharonian06_2356}.
\\
Redshift references:\hfill\break
\tabref{a}\cite{Jones09},
\tabref{b}\cite{Pita14},
\tabref{c}\cite{Sbarufatti05},
\tabref{d}\cite{Rines03},
\tabref{e}\cite{Furniss13},
\tabref{f}\cite{Woo05},
\tabref{g}\cite{Falco99},
\tabref{h}\cite{Donato01},
\tabref{i}\cite{Liu02},
\tabref{j}\cite{Craig97},
\tabref{k}\cite{Falomo99},
\tabref{l}\cite{Shaw13},
\tabref{m}\cite{Paturel02},
\tabref{n}\cite{Massaro09},
\tabref{o}\cite{Rector03},
\tabref{p}\cite{Adelman-McCarthy09},
\tabref{q}\cite{Giommi05},
\tabref{r}\cite{Snellen02},
\tabref{s}\cite{Albert07},
\tabref{t}\cite{Lavaux11},
\tabref{u}\cite{Villforth09},
\tabref{v}\cite{Rines03},
\tabref{w}\cite{Healey2008},
\tabref{x}\cite{Veron10},
\tabref{y}\cite{Fey04},
\tabref{z}\cite{Shaw2012},
\tabref{aa}\cite{Aliu14_1553},
\tabref{ab}\cite{Urry00},
\tabref{ac}\cite{Laurent-Muehleisen99},
\tabref{ad}\cite{Cerruti11},
\tabref{ae}\cite{Berger11},
\tabref{af}\cite{Ganguly13},
\tabref{ag}\cite{Colless01}.

\end{table}
\end{landscape}

\clearpage
\newpage
\begin{table*}  

\caption[]{The comparison of the  Galactic column density values.  (1) The object
name, (2)-(5) the \nh\ values provided by \cite{Kalberla05},
\cite{Dickey90},\cite{Willingale13}, respectively,  (6)-(8) the free \nh\ value
obtained with the \lp, the \po\ and the \bk\ model fitted. The table is available in the online material only.}
\label{table_nh}

\centering

\begin{tabular}{rccccccc}

\hline
 Object & N$_{H}^{LAB}$ & N$_{H}^{DL}$& N$_{H}^{H_2}$  &  N$_{H}^{Will}$ & N$_{H}^{LP,\mathit{free}}$ & N$_{H}^{PL,\mathit{free}}$ & N$_{H}^{BP ,\mathit{free}}$ \\
        &[10$^{20}$\,cm$^{-2}$]&[10$^{20}$\,cm$^{-2}$]&[10$^{20}$\,cm$^{-2}$]&[10$^{20}$\,cm$^{-2}$] &[10$^{20}$\,cm$^{-2}$]  & [10$^{20}$\,cm$^{-2}$] &  [10$^{20}$\,cm$^{-2}$]     \\
    (1)    & (2) & (3) & (4) & (5) & (6) & (7) & (8)\\
\hline
 
SHBL\,J001355.9$-$185406 & 2.13  & 2.09  & 0.17  & 2.30  && \\
KUV\,00311$-$1938        & 1.67  & 1.59  & 0.10  & 1.77  && \\
1ES\,0033$+$595          & 41.80 & 42.70 & 14.40 & 56.20 & 59.36$\pm$0.42 & 68.80$\pm$0.19 \\
RGB\,J0136$+$391         & 6.00  & 5.96  & 1.30  & 7.30  &12.0$\pm$1.5& 11.10$\pm$0.53\\
RGB\,J0152$+$017         & 2.68  & 2.89  & 0.23  & 2.91  && \\
\dots                    & \dots & \dots & \dots &\dots  & \dots          & \dots          & \dots \\

\hline

\end{tabular}
\end{table*}


\begin{table*}

\caption{Fit parameters for the log-parabola fits.  (1) The  object name, (2)
source of \nh\ value, (3)-(5) the $\alpha$, $\beta$ parameters and the
normalization for the \lp\ fit defined in Sect.~\ref{sample},  (6) the reduced
$\chi^2$ value and the number of degrees of freedom.}

\label{table_fitpar_lp} 

\centering

\begin{tabular}{r|c|c|c|c|c}

\hline
Object & N$_H$& $\alpha$ & $\beta$ & $N_l$ & $\chi_{red}^2$/n$_{d.o.f}$ \\
&  &  &  &10$^{-2}$\,cm$^{-2}$\,s$^{-1}$\,keV$^{-1}$ &  \\
(1)    & (2) & (3) & (4) & (5) & (6)  \\
\hline

1ES\,0033$+$595                    & LAB  & 0.93$\pm$0.05 & 0.66$\pm$0.07    & 2.277$\pm$0.044 & 1.048/366\\
                                   & DL   & 0.97$\pm$0.05 & 0.93$\pm$0.07    & 2.327$\pm$0.045 & 1.043/366\\
                                   & Will & 1.49$\pm$0.05 & 0.47$\pm$0.07    & 3.201$\pm$0.065 & 0.999/366\\
                                   & free & 1.60$\pm$0.16 & 0.36$\pm$0.15    & 3.443$\pm$0.338 & 1.001/365\\
\\
RGB\,J0136$+$391                   & LAB  & 2.07$\pm$0.02 & 0.41$\pm$0.04    & 0.718$\pm$0.008 & 0.922/259\\
                                   & DL   & 2.07$\pm$0.02 & 0.41$\pm$0.04    & 0.717$\pm$0.008 & 0.924/259\\
                                   & Will & 2.16$\pm$0.02 & 0.30$\pm$0.04    & 0.744$\pm$0.008 & 0.900/259\\
                                   & free & 2.47$\pm$0.10 & $-$0.07$\pm$0.12 & 0.848$\pm$0.036 & 0.846/258\\
\\                                  
3C\,66A                            & LAB  & 2.45$\pm$0.02 & $-$0.19$\pm$0.04 & 0.136$\pm$0.002 & 0.989/261\\
                                   & DL   & 2.53$\pm$0.02 & $-$0.29$\pm$0.05 & 0.140$\pm$0.002 & 0.960/261\\
                                   & Will & 2.60$\pm$0.02 & $-$0.37$\pm$0.04 & 0.143$\pm$0.002 & 0.943/261\\
                                   & free & 2.76$\pm$0.08 & $-$0.57$\pm$0.10 & 0.153$\pm$0.005 & 0.929/260\\
\\
1ES\,0229$+$200                    & LAB  & 1.41$\pm$0.03 & 0.38$\pm$0.05    & 0.381$\pm$0.006 & 1.162/265\\
                                   & DL   & 1.48$\pm$0.03 & 0.30$\pm$0.05    & 0.384$\pm$0.006 & 1.164/265\\
                                   & Will & 1.63$\pm$0.03 & 0.14$\pm$0.05    & 0.424$\pm$0.006 & 1.178/265\\
                                   & free & 1.40$\pm$0.12 & 0.39$\pm$0.14    & 0.379$\pm$0.022 & 1.166/264\\

\\
PKS\,0447$-$439                    & LAB  & 2.49$\pm$0.01 & 0.27$\pm$0.03    & 0.654$\pm$0.006 & 1.177/273\\
                                   & DL   & 2.55$\pm$0.01 & 0.19$\pm$0.03    & 0.697$\pm$0.006 & 1.154/273\\
                                   & Will & 2.49$\pm$0.01 & 0.27$\pm$0.03    & 0.657$\pm$0.006 & 1.175/273\\
                                   & free & 2.61$\pm$0.04 & 0.10$\pm$0.07    & 0.683$\pm$0.011 & 1.151/272\\
\\
1ES\,0502$+$675                    & LAB  & 1.89$\pm$0.02 & 0.51$\pm$0.03    & 1.315$\pm$0.011 & 1.104/348\\
                                   & DL   & 1.90$\pm$0.02 & 0.49$\pm$0.03    & 1.322$\pm$0.011 & 1.110/348\\
                                   & Will & 2.21$\pm$0.02 & 0.13$\pm$0.03    & 1.517$\pm$0.012 & 1.032/348\\
                                   & free & 2.28$\pm$0.08 & 0.06$\pm$0.10    & 1.564$\pm$0.060 & 1.033/347\\
\\
PKS\,0548$-$322                    & LAB  & 1.73$\pm$0.01 & 0.17$\pm$0.02    & 1.028$\pm$0.006 & 0.956/529\\
                                   & DL   & 1.71$\pm$0.01 & 0.20$\pm$0.02    & 1.016$\pm$0.006 & 0.955/529\\
                                   & Will & 1.75$\pm$0.01 & 0.15$\pm$0.02    & 1.037$\pm$0.006 & 0.959/529\\
                                   & free & 1.70$\pm$0.04 & 0.21$\pm$0.04    & 1.015$\pm$0.016 & 0.957/528\\
\\
1ES\,0647$+$250                    & LAB  & 2.01$\pm$0.12 & 0.45$\pm$0.03    & 0.966$\pm$0.006 & 1.139/426\\
                                   & DL   & 2.07$\pm$0.01 & 0.38$\pm$0.03    & 0.992$\pm$0.006 & 1.128/426\\
                                   & Will & 2.26$\pm$0.01 & 0.17$\pm$0.03    & 1.079$\pm$0.006 & 1.128/426\\
                                   & free & 2.16$\pm$0.06 & 0.27$\pm$0.05    & 1.035$\pm$0.003 & 1.125/425\\
\\
S5\,0716$+$714                     & LAB  & 2.07$\pm$0.01 & $-$0.15$\pm$0.02 & 0.116$\pm$0.001 & 1.029/434\\
                                   & DL   & 2.11$\pm$0.01 & $-$0.20$\pm$0.02 & 0.230$\pm$0.002 & 1.024/434\\
                                   & Will & 2.10$\pm$0.01 & $-$0.19$\pm$0.02 & 0.230$\pm$0.002 & 1.025/434\\
                                   & free & 2.12$\pm$0.04 & $-$0.22$\pm$0.05 & 0.231$\pm$0.004 & 1.026/433\\
\\
1ES\,0806$+$524                    & LAB  & 2.37$\pm$0.01 & 0.29$\pm$0.03    & 0.414$\pm$0.003 & 0.994/284\\
                                   & DL   & 2.39$\pm$0.01 & 0.27$\pm$0.03    & 0.418$\pm$0.003 & 0.990/284\\
                                   & Will & 2.42$\pm$0.01 & 0.23$\pm$0.03    & 0.422$\pm$0.004 & 0.985/284\\
                                   & free & 2.46$\pm$0.06 & 0.18$\pm$0.08    & 0.429$\pm$0.008 & 0.987/283\\
\\
1ES\,1011$+$496                    & LAB  & 2.25$\pm$0.01 & 0.24$\pm$0.02    & 0.938$\pm$0.006 & 1.268/360\\
                                   & DL   & 2.24$\pm$0.01 & 0.24$\pm$0.02    & 0.937$\pm$0.006 & 1.269/360\\
                                   & Will & 2.25$\pm$0.01 & 0.23$\pm$0.02    & 0.938$\pm$0.006 & 1.267/360\\
                                   & free & 2.32$\pm$0.04 & 0.15$\pm$0.05    & 0.961$\pm$0.013 & 1.260/359\\
\\
1ES\,1101$-$232                    & LAB  & 2.02$\pm$0.02 & 0.11$\pm$0.04    & 1.910$\pm$0.022 & 1.034/269\\
                                   & DL   & 2.03$\pm$0.02 & 0.10$\pm$0.04    & 1.916$\pm$0.022 & 1.035/269\\
                                   & Will & 2.11$\pm$0.02 & 0.23$\pm$0.04    & 1.987$\pm$0.023 & 1.053/269\\
                                   & free & 1.89$\pm$0.08 & 0.25$\pm$0.10    & 1.812$\pm$0.061 & 1.028/268\\

\hline
\end{tabular}
\end{table*}


\clearpage
\newpage

\begin{table*}

\contcaption{}

\centering

\begin{tabular}{r|c|c|c|c|c}

\hline
(1)    & (2) & (3) & (4) & (5) & (6)  \\
\hline
Mrk\,180                           & LAB  & 2.09$\pm$0.01 & 0.25$\pm$0.03    & 0.849$\pm$0.006 & 1.012/339\\
                                   & DL   & 2.11$\pm$0.01 & 0.23$\pm$0.03    & 0.854$\pm$0.006 & 1.005/339\\
                                   & Will & 2.10$\pm$0.01 & 0.25$\pm$0.03    & 0.850$\pm$0.006 & 1.011/339\\
                                   & free & 2.21$\pm$0.04 & 0.11$\pm$0.06    & 0.885$\pm$0.015 & 0.992/338\\
\\
RX\,J1136.5$+$6737                 & LAB  & 1.65$\pm$0.02 & 0.23$\pm$0.04    & 0.417$\pm$0.005 & 1.041/271\\
                                   & DL   & 1.68$\pm$0.02 & 0.21$\pm$0.04    & 0.421$\pm$0.005 & 1.040/271\\
                                   & Will & 1.66$\pm$0.02 & 0.23$\pm$0.04    & 0.418$\pm$0.004 & 1.041/271\\
                                   & free & 1.71$\pm$0.08 & 0.17$\pm$0.10    & 0.427$\pm$0.014 & 1.043/270\\
\\
1ES\,1218$+$304                    & LAB  & 2.04$\pm$0.02 & 0.13$\pm$0.04    & 0.927$\pm$0.001 & 1.198/287\\
                                   & DL   & 2.02$\mp$0.02 & 0.15$\pm$0.04    & 0.921$\pm$0.001 & 1.195/287\\
                                   & Will & 2.05$\pm$0.02 & 0.12$\pm$0.04    & 0.930$\pm$0.001 & 1.199/287\\
                                   & free & 1.96$\pm$0.06 & 0.22$\pm$0.08    & 0.902$\pm$0.022 & 1.196/286\\
\\
W\,Comae                           & LAB  & 2.44$\pm$0.01 & $-$0.24$\pm$0.04 & 0.065$\pm$0.001 & 1.015/253\\
                                   & DL   & 2.43$\pm$0.01 & $-$0.21$\pm$0.04 & 0.065$\pm$0.001 & 1.042/253\\
                                   & Will & 2.45$\pm$0.01 & $-$0.25$\pm$0.04 & 0.065$\pm$0.001 & 1.010/253\\
                                   & free & 2.63$\pm$0.06 & $-$0.48$\pm$0.08 & 0.070$\pm$0.002 & 0.968/252\\
\\
MS\,1221.8$+$2452                  & LAB  & 2.18$\pm$0.02 & 0.39$\pm$0.04    & 0.395$\pm$0.005 & 1.143/240\\
                                   & DL   & 2.18$\pm$0.02 & 0.39$\pm$0.04    & 0.396$\pm$0.004 & 1.143/240\\
                                   & Will & 2.19$\pm$0.02 & 0.38$\pm$0.04    & 0.397$\pm$0.005 & 1.139/240\\
                                   & free & 2.38$\pm$0.07 & 0.13$\pm$0.10    & 0.426$\pm$0.012 & 1.110/239\\
\\
4C\,$+$21.35                       & LAB  & 1.87$\pm$0.01 & $-$0.48$\pm$0.03 & 0.064$\pm$0.001 & 1.032/374\\
                                   & DL   & 1.88$\pm$0.01 & $-$0.49$\pm$0.03 & 0.064$\pm$0.001 & 1.035/374\\
                                   & Will & 1.88$\pm$0.01 & $-$0.49$\pm$0.03 & 0.064$\pm$0.001 & 1.036/374\\
                                   & free & 1.75$\pm$0.05 & $-$0.35$\pm$0.06 & 0.061$\pm$0.001 & 1.018/373\\
\\
3C\,279                            & LAB  & 1.48$\pm$0.01 & 0.16$\pm$0.02    & 0.245$\pm$0.001 & 1.108/607\\
                                   & DL   & 1.49$\pm$0.01 & 0.14$\pm$0.02    & 0.246$\pm$0.001 & 1.110/607\\
                                   & Will & 1.49$\pm$0.01 & 0.14$\pm$0.02    & 0.246$\pm$0.001 & 1.108/607\\
                                   & free & 1.45$\pm$0.03 & 0.19$\pm$0.04    & 0.242$\pm$0.003 & 1.108/606\\
\\
PKS\,1424$+$240                    & LAB  & 2.48$\pm$0.01 & 0.34$\pm$0.03    & 0.261$\pm$0.002 & 1.081/267\\
                                   & DL   & 2.47$\pm$0.01 & 0.36$\pm$0.03    & 0.260$\pm$0.003 & 1.093/267\\
                                   & Will & 2.51$\pm$0.01 & 0.30$\pm$0.03    & 0.264$\pm$0.002 & 1.059/267\\
                                   & free & 2.76$\pm$0.05 & $-$0.04$\pm$0.07 & 0.288$\pm$0.006 & 0.968/266\\
\\
H\,1426$+$428                      & LAB  & 1.84$\pm$0.01 & 0.10$\pm$0.01    & 1.423$\pm$0.006 & 1.245/583\\
                                   & DL   & 1.86$\pm$0.01 & 0.18$\pm$0.01    & 1.434$\pm$0.006 & 1.246/583\\
                                   & Will & 1.84$\pm$0.01 & 0.20$\pm$0.01    & 1.424$\pm$0.006 & 1.245/583\\
                                   & free & 1.84$\pm$0.02 & 0.19$\pm$0.03    & 1.425$\pm$0.014 & 1.247/582\\
\\
PKS\,1510$-$089                    & LAB  & 1.39$\pm$0.01 & $-$0.10$\pm$0.02 & 0.107$\pm$0.001 & 1.046/660\\
                                   & DL   & 1.44$\pm$0.01 & $-$0.15$\pm$0.02 & 0.110$\pm$0.001 & 1.061/660\\
                                   & Will & 1.54$\pm$0.01 & $-$0.24$\pm$0.02 & 0.115$\pm$0.001 & 1.103/660\\
                                   & free & 1.28$\pm$0.04 & 0.01$\pm$0.04    & 0.101$\pm$0.002 & 1.033/659\\
\\
PG\,1553$+$113                     & LAB  & 2.25$\pm$0.01 & 0.13$\pm$0.02    & 1.046$\pm$0.006 & 1.299/432\\
                                   & DL   & 2.25$\pm$0.01 & 0.13$\pm$0.02    & 1.046$\pm$0.006 & 1.298/432\\
                                   & Will & 2.30$\pm$0.01 & 0.07$\pm$0.02    & 1.065$\pm$0.006 & 1.288/432\\
                                   & free & 2.33$\pm$0.03 & 0.03$\pm$0.05    & 1.076$\pm$0.015 & 1.290/431\\
\\
1ES\,1959$+$650                    & LAB  & 1.91$\pm$0.01 & 0.40$\pm$0.02    & 7.301$\pm$0.042 & 1.263/466\\
                                   & DL   & 1.94$\pm$0.01 & 0.36$\pm$0.02    & 7.400$\pm$0.042 & 1.230/466\\
                                   & Will & 2.29$\pm$0.01 & $-$0.01$\pm$0.02 & 8.669$\pm$0.049 & 1.331/466\\
                                   & free & 2.00$\pm$0.05 & 0.30$\pm$0.06    & 7.586$\pm$0.177 & 1.260/465\\
\hline
\end{tabular}
\end{table*}


\clearpage
\newpage

\begin{table*}

\contcaption{}

\centering

\begin{tabular}{r|c|c|c|c|c}

\hline
(1)    & (2) & (3) & (4) & (5) & (6)  \\
\hline
PKS\,2005$-$489                    & LAB  & 2.47$\pm$0.01 & $-$0.05$\pm$0.03 & 0.920$\pm$0.007 & 0.984/326\\
                                   & DL   & 2.55$\pm$0.01 & $-$0.14$\pm$0.03 & 0.947$\pm$0.007 & 0.996/326\\
                                   & Will & 2.53$\pm$0.01 & $-$0.11$\pm$0.03 & 0.937$\pm$0.007 & 0.989/326\\
                                   & free & 2.47$\pm$0.04 & $-$0.04$\pm$0.06 & 0.919$\pm$0.016 & 0.987/325\\
\\
PKS\,2155$-$304                    & LAB  & 2.38$\pm$0.02 & 0.23$\pm$0.05    & 3.834$\pm$0.050 & 1.151/208\\
                                   & DL   & 2.39$\pm$0.02 & 0.21$\pm$0.05    & 3.852$\pm$0.051 & 1.149/208\\
                                   & Will & 2.38$\pm$0.02 & 0.22$\pm$0.05    & 3.844$\pm$0.051 & 1.150/208\\
                                   & free & 2.45$\pm$0.07 & 0.13$\pm$0.11    & 3.926$\pm$0.106 & 1.151/207\\
\\
BL\,Lacertae                       & LAB  & 1.23$\pm$0.02 & 0.48$\pm$0.02    & 0.266$\pm$0.002 & 1.164/615\\
                                   & DL   & 1.40$\pm$0.02 & 0.31$\pm$0.02    & 0.292$\pm$0.002 & 1.122/615\\
                                   & Will & 1.82$\pm$0.02 & $-$0.10$\pm$0.02 & 0.369$\pm$0.002 & 1.146/615\\
                                   & free & 1.55$\pm$0.06 & 0.16$\pm$ 0.06   & 0.318$\pm$0.010 & 1.111/614\\
\\
B3\,2247$+$381                     & LAB  & 2.03$\pm$0.02 & 0.52$\pm$0.04    & 0.375$\pm$0.004 & 1.036/282\\
                                   & DL   & 2.15$\pm$0.02 & 0.36$\pm$0.04    & 0.396$\pm$0.004 & 1.021/282\\
                                   & Will & 2.35$\pm$0.02 & 0.13$\pm$0.04    & 0.433$\pm$0.004 & 1.023/282\\
                                   & free & 2.24$\pm$0.10 & 0.27$\pm$0.12    & 0.411$\pm$0.019 & 1.022/281\\
\\
1ES\,2344$+$514                    & LAB  & 1.86$\pm$0.01 & 0.29$\pm$0.01    & 0.528$\pm$0.003 & 1.080/531\\
                                   & DL   & 1.96$\pm$0.01 & 0.18$\pm$0.02    & 0.552$\pm$0.003 & 1.093/531\\
                                   & Will & 2.40$\pm$0.01 & $-$0.26$\pm$0.02 & 0.685$\pm$0.004 & 1.308/531\\
                                   & free & 1.81$\pm$0.05 & 0.33$\pm$0.06    & 0.514$\pm$0.013 & 1.080/530\\

\hline
\end{tabular}
\end{table*}

\clearpage
\newpage

\begin{table*}

\caption{Fit parameters for the power-law fits.  (1) The  object name, (2) source of \nh\ value, (3)-(4) the spectral index and the normalization for the \po\
fit defined in Sect.~\ref{sample},   (5) the reduced $\chi^2$ value and the
number of degrees of freedom, (6) the test statistics value for F-test, (7)
probability, (8) preferred spectral model: PO - for the \po\ and LP - for the  \lp\ model. The table is available in the  online material only.}

\label{table_fitpar_po}

\centering

\begin{tabular}{rccccccc}

\hline
Object & N$_H$& $\Gamma$ & $N_p$ & $\chi_{red}^2$/n$_{d.o.f}$ &  F &  p-value & Model  \\
&  &    &10$^{-2}$\,cm$^{-2}$\,s$^{-1}$\,keV$^{-1}$ & &   & &\\
(1)    & (2) & (3) & (4) & (5) & (6)  & (7)  & (8) \\
\hline

1ES\,0033$+$595    & LAB  & 1.62$\pm$0.01 & 2.581$\pm$0.037 & 1.800/367& 264& 1$\cdot10^{-16}$ &  LP  \\
                   & DL   & 1.64$\pm$0.01 & 2.635$\pm$0.038 & 1.736/367& 244& 1$\cdot10^{-16}$ &  LP  \\
                   & Will & 1.84$\pm$0.02 & 3.476$\pm$0.052 & 1.143/367&  53& 1$\cdot10^{-12}$ &  LP  \\
                   & free & 1.99$\pm$0.03 & 4.322$\pm$0.151 & 1.013/366&   5& 0.02             &  PO  \\
\\
RGB\,J0136$+$391   & LAB  & 2.17$\pm$0.01 & 0.671$\pm$0.006 & 1.321/260& 114& 1$\cdot10^{-16}$ & LP  \\
                   & DL   & 2.17$\pm$0.01 & 0.670$\pm$0.006 & 1.329/260& 115& 1$\cdot10^{-16}$ & LP  \\
                   & Will & 2.23$\pm$0.01 & 0.711$\pm$0.007 & 1.094/260&  57& 1$\cdot10^{-13}$ & LP  \\
                   & free & 2.42$\pm$0.03 & 0.830$\pm$0.018 & 0.845/259& 0.7& 0.4              & PO  \\

\dots              & \dots & \dots        & \dots           & \dots    & \dots & \dots & \dots\\
\hline
\end{tabular}
\end{table*}

\begin{table*}

\caption{Fit parameters for the \bk\ fits.  (1) The object
name, (2) source of \nh\ value, (3)-(6) the spectral indices, normalization and the 
break energy for the \bk\ fit defined in Sect.~\ref{sample}, (7) the reduced
$\chi^2$ value and the number of degrees of freedom. The table is available in the  online material only.}

\label{table_fitpar_bk}

\begin{tabular}{rcccccc}
\hline
Object & N$_H$& $\Gamma_1$ & $\Gamma_2$ & $N_b$ & $E_{b}$ & $\chi_{red}^2$/n$_{d.o.f}$ \\
       &     &     &     &10$^{-2}$\,cm$^{-2}$\,s$^{-1}$\,keV$^{-1}$  & keV & \\
(1)    & (2) & (3) & (4) & (5) & (6) & (7)  \\

\hline
3C\,66A            & LAB  &2.462$\pm$0.017   &  1.746$\pm$0.170   &  0.139$\pm$0.001 &  3.383$\pm$0.328  & 0.949/260\\
                   & DL   &2.540$\pm$0.018   &  1.794$\pm$0.140   &  0.145$\pm$0.001 &  3.064$\pm$0.261  & 0.961/260\\
                   & Will &2.615$\pm$0.020   &  1.936$\pm$0.104   &  0.150$\pm$0.001 &  2.568$\pm$0.213  & 0.985/260\\
                   & free &2.470$\pm$0.040   &  1.739$\pm$0.168   &  0.140$\pm$0.001 &  3.372$\pm$0.340  & 0.952/259\\
\\
S5\,0716$+$714     & LAB  & 2.168$\pm$0.024  &  1.937$\pm$0.016   &  0.221$\pm$0.003 &  1.094$\pm$0.100 & 0.986/433\\
                   & DL   & 2.240$\pm$0.026  &  1.942$\pm$0.015   &  0.223$\pm$0.003 &  1.087$\pm$0.075 & 0.967/433\\
                   & Will & 2.213$\pm$0.025  &  1.940$\pm$0.015   &  0.223$\pm$0.003 &  1.092$\pm$0.082 & 0.973/433\\
                   & free & 2.511$\pm$0.085  &  1.965$\pm$0.017   &  0.231$\pm$0.002 &  1.046$\pm$0.042 & 0.942/432\\
\dots              & \dots&\dots             &  \dots             &  \dots           &  \dots            & \dots \\

\hline
\end{tabular}
\end{table*}

\begin{table*}

\caption[]{The summary of time-resolved spectral analysis.  The table summarize
the spectral parameters for the \bk\ fit to \xrt\ data for the selected
blazars. The time intervals are shown in Fig.\ref{lc_3c66a}-\ref{lc_4c2135},
while the corresponding spectral fits are presented in Fig.~\ref{seds_breaks}.
The following columns show: (1) the name of the object, (2) the selected
interval, (3) the normalization , (4)-(5) the spectral indices, (6) the break
energy (7) the value of reduced $\chi^2$ and the number of degree of freedom.
The fits included in this Table are plotted in Fig.~\ref{seds_breaks}.} 

\centering

\begin{tabular}{r|c|c|c|c|c|c}
\hline
Object & Interval & $N_{b}$  & $\Gamma_1$ & $\Gamma_2$ & $E_{b}$ & $\chi_{red}^2$/n$_{d.o.f}$ \\
 &  & 10$^{-3}$\,cm$^{-2}$\,s$^{-1}$\,keV$^{-1}$ &  &  & keV &  \\
(1) & (2) & (3) & (4) & (5) & (6) & (7) \\
\hline
3C\,66A & A  & 1.173$\pm$0.023  &  2.420$\pm$0.042 & 2.142$\pm$0.044 & 1.37$\pm$0.21 &0.998/219    \\
3C\,66A & B  & 3.158$\pm$0.052  &  2.653$\pm$0.029 & 1.4$\pm$1.8 & 5.2$\pm$1.8& 0.883/139   \\
3C\,66A & C  & 2.62$\pm$0.41  &  1.2$\pm$1.8 & 2.50$\pm$0.33 & 0.46$\pm$0.11 &  0.861/136  \\
\\
S5\,0716+714 & A  & 1.767$\pm$0.072  &  2.911$\pm$0.071 & 2.052$\pm$0.073 &  1.18$\pm$0.11 & 0.976/107 \\
S5\,0716+714 & B  & 2.285$\pm$0.028  &  2.511$\pm$0.024 & 2.035$\pm$0.030 &  1.361$\pm$0.083 & 0.925/310 \\
S5\,0716+714 & C &  2.220$\pm$0.029  &  2.258$\pm$0.025 & 1.932$\pm$0.018 &  1.153$\pm$0.079 & 1.009/408 \\
S5\,0716+714 & D  & 1.872$\pm$0.098  &  2.122$\pm$0.090 & 1.976$\pm$0.037 &  0.92$\pm$0.39 & 0.841/188   \\
\\
W~Comae &  A & 0.855$\pm$0.079  &  2.550$\pm$0.016 & 1.84$\pm$0.20 &  4.07$\pm$0.46 & 0.956/163  \\
W~Comae & B  & 0.263$\pm$0.014  &  2.18$\pm$0.10 & 1.738$\pm$0.067 &  1.18$\pm$0.23 & 0.814/298  \\
\\
4C\,$+$21.35 & A  & 0.556$\pm$0.015  &  2.071$\pm$0.054 & 1.437$\pm$0.032 &  1.180$\pm$0.087 & 0.835/253 \\
4C\,$+$21.35 & B  & 0.393$\pm$0.028  &  1.71$\pm$0.13   & 1.366$\pm$0.038 &  0.97$\pm$0.22 & 1.125/136 \\
4C\,$+$21.35 & C  & 0.915$\pm$0.025  &  2.340$\pm$0.052 & 1.454$\pm$0.036 &1.209$\pm$0.063 & 0.877/230  \\
\\
BL Lacertae & A  & 3.132$\pm$0.035  &  2.346$\pm$0.039 & 1.941$\pm$0.014 &  1.237$\pm$0.062 & 1.129/523 \\
BL Lacertae & B  & 4.176$\pm$0.080  &  1.947$\pm$0.066 & 1.703$\pm$0.012 &  1.10$\pm$0.12 & 1.143/533  \\
BL Lacertae & C  & 2.245$\pm$0.053  &  2.111$\pm$0.078 & 1.782$\pm$0.016 &  1.10$\pm$0.11 & 0.936/430 \\
\hline
\end{tabular}

\label{table_fits}
\end{table*}

\clearpage
\newpage

\section*{Acknowledgements}
A.W. acknowledges support by Polish Ministry of Science and Higher Education in Mobility Plus Program and  Polish National Science Center for supporting this work  through grant  DEC$-$2011/03/N/ST9/01867.
Support from the German Ministry for Education and
Research (BMBF) through the Verbundforschung Astroteilchenphysik grant 05A11VH2 is gratefully acknowledged.
This research was supported in part by PLGrid Infrastructure.

\bibliographystyle{mn2e_williams}
\bibliography{references}

\label{lastpage}
\clearpage
\includepdf[pages=-]{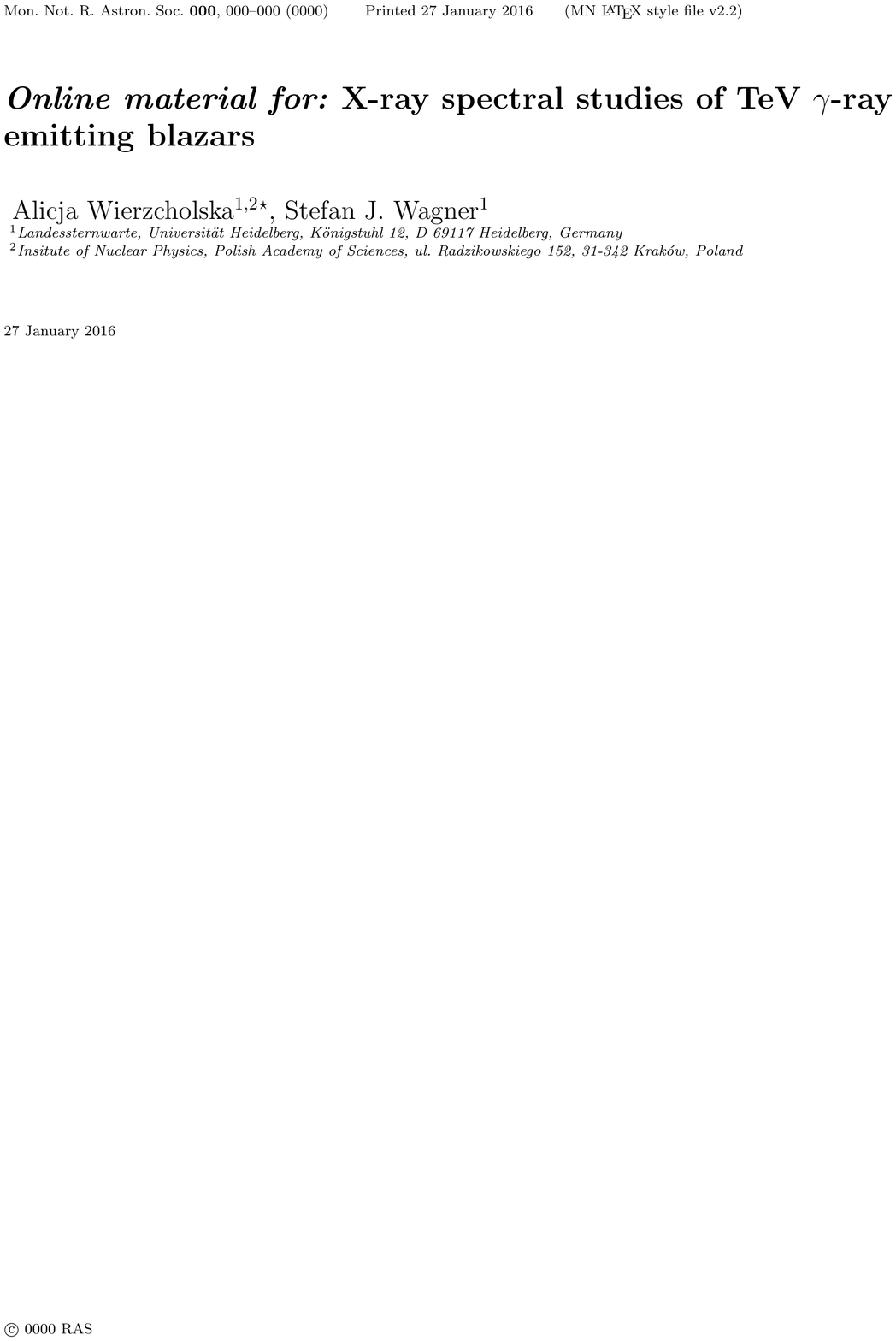}
\end{document}